\documentclass[
  aps,
  prx,
  reprint,
  amsmath,
  amssymb,
  superscriptaddress,
  nofootinbib
]{revtex4-2}

\usepackage{graphicx}

\usepackage{iftex}
\ifPDFTeX
  \usepackage[T1]{fontenc}
  \DeclareUnicodeCharacter{03B1}{\ensuremath{\alpha}}
  \DeclareUnicodeCharacter{03B2}{\ensuremath{\beta}}
  \DeclareUnicodeCharacter{03BC}{\ensuremath{\mu}}
  \DeclareUnicodeCharacter{03C3}{\ensuremath{\sigma}}
  \DeclareUnicodeCharacter{0394}{\ensuremath{\Delta}}
  \DeclareUnicodeCharacter{03B7}{\ensuremath{\eta}}
  \DeclareUnicodeCharacter{03BA}{\ensuremath{\kappa}}
  \DeclareUnicodeCharacter{1D447}{\ensuremath{T}}
  \DeclareUnicodeCharacter{1D6FC}{\ensuremath{\alpha}}
  \DeclareUnicodeCharacter{2009}{\,}
  \DeclareUnicodeCharacter{00B1}{\ensuremath{\pm}}
\else
  \usepackage{newunicodechar}
  \newunicodechar{α}{\ensuremath{\alpha}}
  \newunicodechar{β}{\ensuremath{\beta}}
  \newunicodechar{μ}{\ensuremath{\mu}}
  \newunicodechar{σ}{\ensuremath{\sigma}}
  \newunicodechar{Δ}{\ensuremath{\Delta}}
  \newunicodechar{η}{\ensuremath{\eta}}
  \newunicodechar{κ}{\ensuremath{\kappa}}
\fi

\usepackage{xcolor}
\definecolor{citecol}{HTML}{2A9D8F}
\definecolor{linkcol}{HTML}{2B5E9E}
\definecolor{urlcol}{HTML}{E76F51}
\usepackage[
  colorlinks=true,
  citecolor=citecol,
  linkcolor=linkcol,
  urlcolor=urlcol,
  breaklinks=true,
  pdfusetitle
]{hyperref}

\begin{document}

\title{Autonomous cooling of Kerr-cat qubits to minutes-long bit-flip times}

\author{Sergey Hazanov}
\thanks{These authors contributed equally to this work.}
\affiliation{Department of Condensed Matter Physics, Weizmann Institute of Science, Rehovot, Israel}

\author{Mikhail Svintsov}
\thanks{These authors contributed equally to this work.}
\affiliation{Department of Condensed Matter Physics, Weizmann Institute of Science, Rehovot, Israel}

\author{Uri Goldblatt}
\affiliation{Department of Condensed Matter Physics, Weizmann Institute of Science, Rehovot, Israel}

\author{Barkay Guttel}
\affiliation{Department of Condensed Matter Physics, Weizmann Institute of Science, Rehovot, Israel}

\author{Fabien Lafont}
\thanks{These authors co-supervised the project. Electronic address: \href{mailto:serge.rosenblum@weizmann.ac.il}{serge.rosenblum@weizmann.ac.il}}
\affiliation{Department of Condensed Matter Physics, Weizmann Institute of Science, Rehovot, Israel}

\author{Serge Rosenblum}
\thanks{These authors co-supervised the project. Electronic address: \href{mailto:serge.rosenblum@weizmann.ac.il}{serge.rosenblum@weizmann.ac.il}}
\affiliation{Department of Condensed Matter Physics, Weizmann Institute of Science, Rehovot, Israel}
\affiliation{QuamCore, Herzliya 4673306, Israel}

\date{\today}

\begin{abstract}
Quantum error correction becomes more hardware efficient when physical noise is strongly biased towards one error channel~\cite{aliferis2008, guillaud2019, putterman2025, pappalardo2026}. Cat qubits realise such a bias by encoding information in two opposite coherent states of a superconducting oscillator, so that bit flips are exponentially suppressed with photon number while phase-flip rates increase only linearly~\cite{mirrahimi2014, lescanne2020}. In Kerr-cat qubits, a driven double-well potential confines these states and enables fast coherent control, but excitations can induce interwell
switching, limiting bit-flip times to milliseconds~\cite{frattini2024prx, ding2025, adinolfi2025}. Here we show that the nonlinear interaction generating the confinement can also cool these excitations. When an intra-well transition becomes resonant with a lossy readout mode, excitations are transferred out of the qubit and dissipated, as established by spectroscopy and relaxation measurements. We observe the predicted exponential scaling of Kerr-cat bit-flip times with photon number, reaching 450~s—over five orders of magnitude beyond previous demonstrations. Phase-flip rates remain consistent with intrinsic photon loss at operating points with seconds-long bit-flip times. We further demonstrate fast $Z$ gates without measurable degradation of bit-flip protection and single-shot readout with a bit-flip probability below $10^{-6}$ per measurement. 
Autonomous cooling thus brings minutes-long bit-flip protection to Kerr-cat qubits while retaining fast coherent control, without additional cooling drives or circuit elements.

\end{abstract}

\maketitle

\addtocontents{toc}{\protect\suppltochide}

\begin{figure*}[t]
  \centering
  \includegraphics[width=\textwidth]{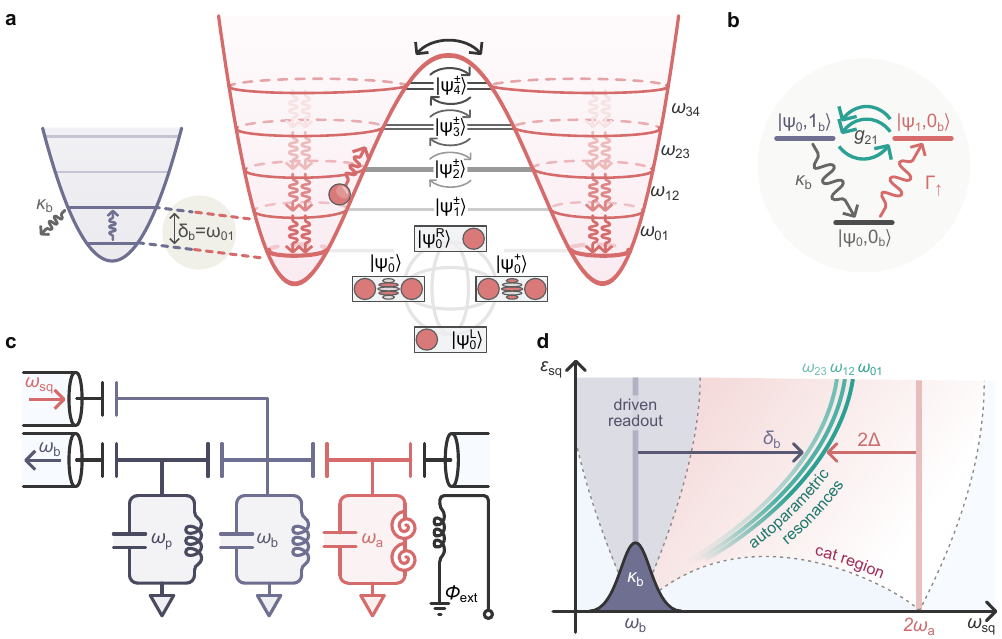}
  \caption{\textbf{Autonomous cooling of a Kerr-cat qubit through its readout circuitry.}
  \textbf{a,} Schematic of the Kerr-cat double-well potential coupled to a lossy readout mode. The well ground states, $|\psi_0^{L}\rangle$ and $|\psi_0^{R}\rangle$, encode the logical $Z$ basis of the cat qubit. Noise populates excited doublets $|\psi_n^{\pm}\rangle$, separated by intra-well transition frequencies $\omega_{n-1,n}$, from which interwell tunnelling becomes more likely (curved arrows).
  When the drive-readout detuning $\delta_\mathrm{b}=\omega_{\mathrm{sq}}-\omega_\mathrm{b}$ matches $\omega_{01}$, a nonlinear exchange interaction transfers intra-well excitations (red wavy arrows) into readout photons (blue wavy arrow), which decay at rate $\kappa_\mathrm{b}$ (grey wavy arrow). \textbf{b,} Cooling cycle for the first excited manifold.
  Heating at rate $\Gamma_\uparrow$ excites the oscillator. The exchange interaction $g_{21}$ transfers the excitation to the readout resonator, whose decay at rate $\kappa_b$ returns the system to the ground manifold.
  \textbf{c,} Device circuit. A SNAIL oscillator with flux-tunable frequency $\omega_a$ is capacitively coupled to a readout resonator of frequency $\omega_b$.
Driving the readout resonator at $\omega_{\mathrm{sq}}$ generates the two-photon squeezing drive through the SNAIL's nonlinearity.
The readout resonator is coupled to the output through a Purcell filter of frequency $\omega_p$, which provides a second cooling channel when $\delta_p=\omega_{\mathrm{sq}}-\omega_p$ matches $\omega_{01}$.
Both cooling channels use the existing readout circuitry, without additional drives or circuit elements.
\textbf{d,} Schematic drive parameter space at fixed flux bias, showing the squeezing-drive amplitude $\varepsilon_{\mathrm{sq}}$ versus frequency $\omega_{\mathrm{sq}}$.
Red shading marks the Kerr-cat regime, while purple shading indicates the region of high readout-resonator population.
Horizontal arrows indicate the detunings $\delta_b$ and $2\Delta=2\omega_a-\omega_{\mathrm{sq}}$.
Teal curves mark the autoparametric cooling resonances $\omega_{n-1,n}=\delta_b$ for the three lowest intra-well transitions.
  }
  \label{fig:concept}
\end{figure*}

Qubits with strongly biased noise can substantially reduce the overhead of quantum error correction. For such qubits, error-correcting codes can be tailored to the dominant
channel, from modified surface codes to a repetition code in the limit of large
bias~\cite{aliferis2008, bonillaataides2021, darmawan2021, guillaud2019, pappalardo2026, putterman2025}. Cat qubits
naturally realise such a bias~\cite{mirrahimi2014, puri2019}. Their codewords are coherent states at opposite points of an
oscillator's phase space, so that a bit flip requires the oscillator to move from one to the other. The bit-flip rate is therefore suppressed exponentially with the mean photon number $\bar n$, whereas the phase-flip rate
grows only linearly~\cite{lescanne2020}.

Two principal approaches have been developed to confine cat qubits. Dissipative cats exchange photon
pairs with a lossy mode~\cite{mirrahimi2014, leghtas2015, reglade2024, paradina2026}, continuously pumping the oscillator towards the code space~\cite{marquet2024}. They have achieved bit-flip times exceeding 100 s~\cite{berdou2023} and have recently been concatenated into a repetition
code~\cite{putterman2025prx, putterman2025}. Kerr cats instead separate the codewords by an energy barrier: a squeezing drive applied to a Kerr-nonlinear
oscillator produces an effective double-well potential whose logical ground-state manifold is separated from excited states by an energy gap~\cite{goto2016, puri2017, grimm2020} (Fig.~\ref{fig:concept}a). This Hamiltonian confinement supports fast, bias-preserving coherent control~\cite{puri2020}, with nonadiabatic gate errors that can be exponentially suppressed with gate time~\cite{gautier2022, xu2022}. Yet Kerr-cat bit-flip times have remained at the millisecond scale~\cite{grimm2020, frattini2024prx, hajr2024,
adinolfi2025}, five orders of magnitude below the lifetimes demonstrated with dissipative cats.

This limitation arises because Hamiltonian confinement, unlike dissipative stabilisation, does not itself remove excitations from the system. Noise can populate excited states within either well, from which switching between the wells becomes much more likely~\cite{gautier2022, frattini2024prx, xu2022, su2025, marthaler2006, carde2026,putterman2022}. The resulting bit-flip protection therefore depends not only on the height of the energy barrier, but also on the population and tunnelling rates of these excited states.
Frequency-selective~\cite{putterman2022} and voltage-based~\cite{masuda2026} dissipation have been proposed to return these excitations to the code space while suppressing additional logical phase-flip errors. The former was demonstrated through an additional driven coupling between the first excited manifold and a lossy readout resonator~\cite{ding2025, adinolfi2025}, extending Kerr-cat bit-flip times to 3.6~ms~\cite{adinolfi2025}. A complementary strategy suppresses the tunnelling itself by detuning the squeezing drive to bring excited doublets into degeneracy~\cite{ruiz2023, venkatraman2024pnas, qing2026}.
Increasing the photon number should further suppress tunnelling~\cite{gautier2022, xu2022}, but at strong driving this improvement is typically interrupted by multiphoton resonances with the readout resonator, which sharply degrade the bit-flip protection~\cite{benhayounekhadraoui2026}.

Here we show that such multiphoton resonances, rather than limiting Kerr-cat protection, can instead provide an autonomous cooling channel, without additional cooling drives or circuit elements. 
The same nonlinear interaction that generates the confining squeezing drive also couples excitations within each well to the lossy readout circuitry. Near resonance, this coupling drains population from multiple excited levels, suppressing the interwell switching responsible for bit-flips.  Importantly, this interaction cools the excited states while leaving the computational manifold essentially unaffected.

We resolve the underlying multiphoton resonances spectroscopically and directly measure the accelerated relaxation of excited states. The bit-flip time is strongly enhanced along the cooling resonances and increases exponentially with the mean photon number $\bar{n}$, reaching 450~s—more than five orders of magnitude beyond previous Kerr-cat qubits. At the same time, the phase-flip rate retains its expected linear dependence on photon number. We further demonstrate that coherent \(Z\) control introduces no measurable degradation of bit-flip protection, while repeated readout induces a bit-flip probability of \(5\times10^{-7}\) per measurement.

\begin{figure*}[!t]
  \centering
  \includegraphics[width=\textwidth]{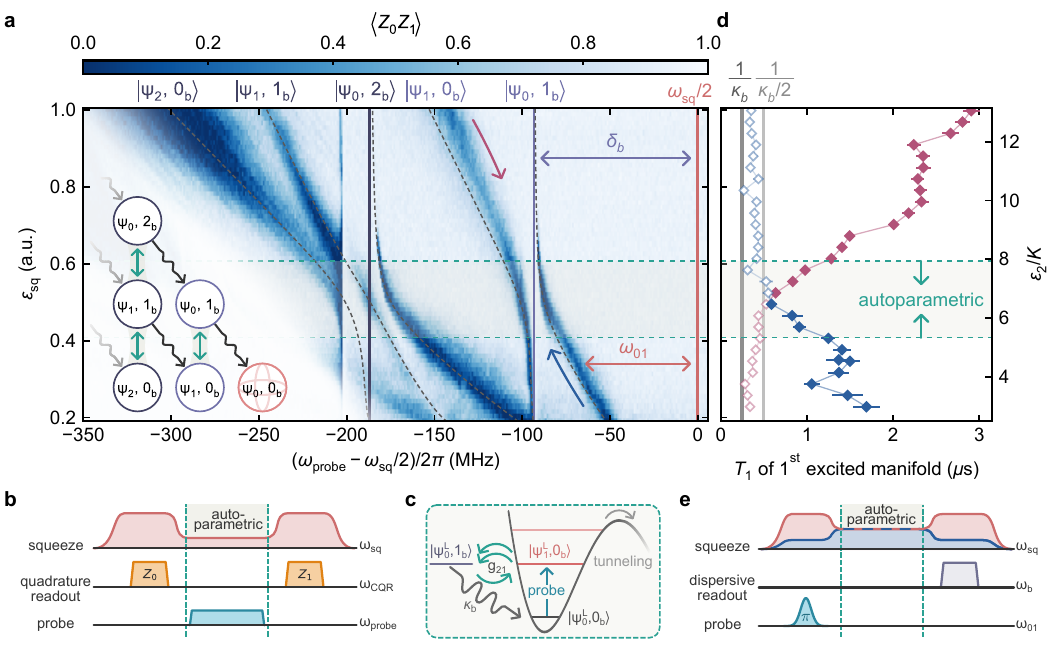}
  \caption{\textbf{Autoparametric hybridisation and cooling of Kerr-cat excitations.}
  \textbf{a,} Excited-state spectroscopy showing $\langle Z_0Z_1\rangle$, the mean product of the two cat-quadrature readouts $Z_0$ and $Z_1$ (\textbf{b}), versus the detuning of a $5$-$\mu$s probe from $\omega_{\mathrm{sq}}/2$ and squeezing-drive amplitude $\varepsilon_{\mathrm{sq}}$. Dashed grey lines show the fitted two-mode model (Methods), with states labelled $|\psi_n, n_b\rangle$, where $n$ denotes the intra-well excitation level and $n_b$ the readout photon number. Increasing the squeezing amplitude shifts intra-well transition frequencies through the readout resonance, producing avoided crossings at $\omega_{01}=\delta_b$ and near $\omega_{02}=2\delta_b$, where $\delta_b=\omega_{\mathrm{sq}}-\omega_b$ (Supplementary Information Sec.~\ref{note:chains}). Vertical lines mark the drive-independent states $|\psi_0,1_b\rangle$ and $|\psi_0,2_b\rangle$, and the reference frequency $\omega_{\mathrm{sq}}/2$. The fit is used to calibrate the effective squeezing amplitude $\varepsilon_2/K$, where $K$ is the oscillator's Kerr nonlinearity. The inset illustrates the
  cooling cascade: successive exchanges (green arrows) transfer intra-well excitations into readout photons, which decay at rate $\kappa_b$ (wavy arrows). Shading marks the autoparametric band. Coloured arrows indicate sweep directions in \textbf{d}. The feature at $-203$~MHz is a spurious mode.
  \textbf{b,} Spectroscopy pulse sequence, with a probe between two cat-quadrature readouts (Methods).
  \textbf{c,} Level scheme showing probe excitation, interwell tunnelling (which reduces $\langle Z_0Z_1\rangle$), and competing relaxation through the readout resonator.
  \textbf{d,} Relaxation time $T_1$ of the first excited manifold versus $\varepsilon_{\mathrm{sq}}$. Pink (blue) markers denote preparation above (below) the autoparametric band; open symbols indicate measurements in which the target amplitude lies on the opposite side of the resonance from the preparation point. Grey vertical lines mark $\kappa_b^{-1}$ and the strong-coupling relaxation time $(\kappa_b/2)^{-1}$.
Error bars are standard errors of the exponential fits.
  \textbf{e,} Pulse sequence for the relaxation measurement in \textbf{d}. A resonant $\pi$ pulse prepares the first excited state, followed by a ramp to the target squeezing amplitude, a variable hold time, and dispersive readout after returning to the preparation point.
  Device parameters are $\delta_b/2\pi=93.5$~MHz and $K/2\pi=1.97$~MHz.}
  \label{fig:spectroscopy}
\end{figure*}

\begin{figure*}[t]
  \centering
  \includegraphics[width=\textwidth]{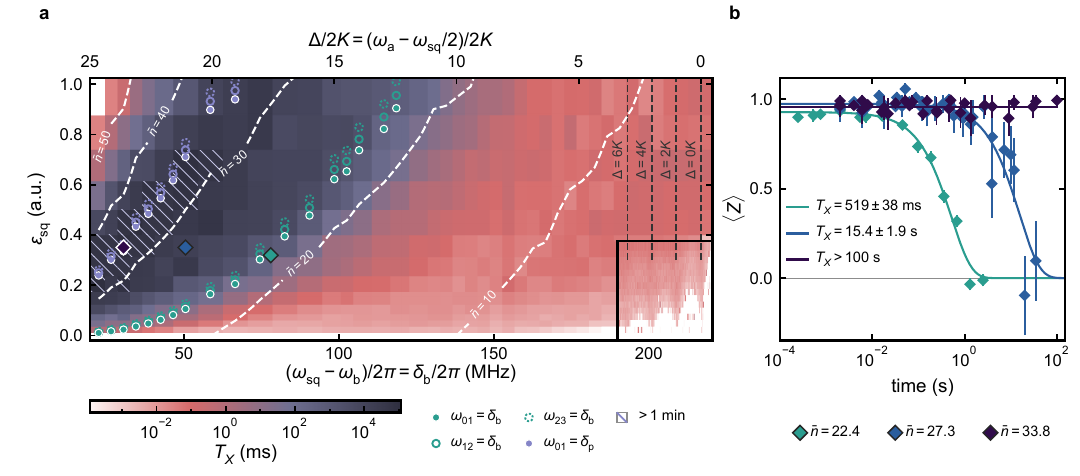}
  \caption{\textbf{Extending Kerr-cat qubit bit-flip times with autonomous cooling.}
  \textbf{a,} Bit-flip time $T_X$ versus squeezing-drive detuning $\omega_{\mathrm{sq}}-\omega_b=\delta_b$ and amplitude $\varepsilon_{\mathrm{sq}}$ at fixed flux bias; the top axis shows the oscillator detuning from half the drive frequency, $\Delta=\omega_a-\omega_{\mathrm{sq}}/2$, in units of $2K$. The inset shows finer sampling of the conventional Kerr-cat regime near $\Delta=0$. Black dashed lines mark the detuning
  degeneracies $\Delta=2mK$. 
  Teal markers indicate resonances of the first three intra-well transitions with the readout resonator
($\omega_{n-1,n}=\delta_b$), with filled, open, and dotted rings distinguishing $\omega_{01}$, $\omega_{12}$, and $\omega_{23}$. Purple markers indicate the resonance of the first transition
with the Purcell filter ($\omega_{01}=\delta_p$), where
$\delta_p=\omega_{\mathrm{sq}}-\omega_p$. White dashed contours mark constant mean photon number $\bar n$. Both the resonance markers and photon-number contours are determined independently from spectroscopy (Fig.~\ref{fig:spectroscopy}a). White cells indicate regions without data. Hatched cells indicate $T_X>1$~min, including lower bounds where no decay was observed. Diamonds indicate the operating points in \textbf{b}:
  on the autoparametric resonance with the readout resonator (teal), between the two resonances (blue), and on the Purcell filter resonance (dark purple).
  \textbf{b,} Bit-flip decay curves at the three operating points. On the autoparametric
  resonance with the readout resonator ($\bar n=22.4\pm1.1$), $T_X=519\pm38$~ms. Between the
  two resonances ($\bar n=27.3\pm1.4$), $T_X=15.4\pm1.9$~s. On the Purcell-filter resonance ($\bar n=33.8\pm1.7$), no decay is observed within the $100$~s measurement window, giving \(T_X>100\) s.
  Uncertainties in $\bar n$ reflect the uncertainty in $K$. Other error bars and quoted uncertainties are one standard error.}
  \label{fig:lifetime}
\end{figure*}

Our device consists of an oscillator with flux-tunable frequency $\omega_a$ (Fig.~\ref{fig:concept}c), incorporating two superconducting nonlinear asymmetric inductive elements (SNAILs)~\cite{frattini2017}. The oscillator is coupled to a lossy readout resonator of frequency $\omega_b$, which connects to the output line through a Purcell filter. The SNAIL's cubic nonlinearity provides a two-to-one exchange between the oscillator and readout resonator, $g_{21}(\hat a^2\hat b^\dagger+\mathrm{h.c.})$, where $\hat a$ and $\hat b$ are the oscillator and readout-resonator annihilation operators and $g_{21}$ is the exchange strength. 
Driving the readout resonator generates two-photon squeezing in the oscillator through this interaction. Together with the oscillator’s Kerr nonlinearity, the squeezing produces the effective double-well potential that confines the cat qubit~\cite{puri2017,grimm2020} (Fig.~\ref{fig:concept}a, Supplementary Information Sec.~\ref{note:hamiltonian}).
The same interaction also provides a route to cooling excitations within each well. 
In the shifted Fock basis~\cite{xu2022, putterman2022}, the oscillator factors into a logical qubit and an intra-well mode, with $\hat a \approx \hat Z\otimes(\alpha + \hat d)$. Here $\hat Z$ is the logical Pauli operator that distinguishes the two wells, $\alpha=\sqrt{\bar{n}}$
is the coherent amplitude of the cat, $\bar{n}$ is its mean photon number, and $\hat d$ is the intra-well annihilation operator. The exchange interaction then contains a beam-splitter term,
\begin{equation}
  2g_{21}\sqrt{\bar{n}}\;\hat I\otimes\big(\hat d\,\hat b^\dagger + \hat d^\dagger\hat b\big),
  \label{eq:beamsplitter}
\end{equation}
coupling the intra-well excitations to the readout resonator while acting as the identity on the logical qubit. Its coupling strength $2g_{21}\sqrt{\bar{n}}$ grows with the mean photon number. Cooling occurs when an intra-well transition of frequency $\omega_{n-1,n}$ matches the detuning $\delta_b=\omega_{\mathrm{sq}}-\omega_b$ between the squeezing drive and the readout resonator (Methods and Supplementary Information Sec.~\ref{note:gap}). At this autoparametric resonance~\cite{marquet2024}, excitations transfer to the readout resonator, where they decay at rate $\kappa_b=2\pi\times0.645$~MHz (Figs.~\ref{fig:concept}b,d). Because the exchange is reversible, thermal photons in the readout resonator can also populate the intra-well excited states, setting a floor on the population that cooling can achieve (Supplementary Information Sec.~\ref{note:floor}).

We probe the predicted exchange interaction between the Kerr oscillator and the readout resonator through spectroscopy of the intra-well excitations (Fig.~\ref{fig:spectroscopy}a). Because the readout resonator also serves as the cooling buffer, cat-quadrature readout is degraded at the autoparametric resonance. We therefore use a dynamical measurement scheme (Fig.~\ref{fig:spectroscopy}b): two cat-quadrature readouts~\cite{grimm2020} are performed at the same squeezing-drive setting, while between them, the squeezing amplitude is ramped to the value under study and a weak probe is applied. Probe-induced transitions reduce $\langle Z_0Z_1\rangle$, the mean product of the two readout outcomes, revealing the excitation spectrum.
As the squeezing drive increases, the first excited state within each well comes into resonance with the ground state plus one readout photon. The exchange interaction hybridises these states, producing an anticrossing (Fig.~\ref{fig:spectroscopy}a). One rung higher, the second excited state hybridises with the first excited state plus one readout photon and the ground state plus two, forming a three-level anticrossing (Fig.~\ref{fig:spectroscopy}a; full spectrum in Supplementary Information Sec.~\ref{note:chains}). The hybridisation is sufficiently strong that the resonances of successive intra-well transitions overlap in squeezing amplitude, forming the autoparametric band shaded in Fig.~\ref{fig:spectroscopy}a (Supplementary Information Sec.~\ref{note:merging}).

To establish that this hybridisation enables cooling, we measure the relaxation of the first excited manifold across the autoparametric resonance (Fig.~\ref{fig:spectroscopy}d,e).
We prepare the excitation on either side of the anticrossing, tune the squeezing drive to the amplitude under study for a variable hold time, and then measure the remaining excited-state population dispersively~\cite{adinolfi2025}.
On resonance, the relaxation time drops to $(\kappa_b/2)^{-1}$, the strong-coupling limit set by the readout linewidth (Supplementary Information Sec.~\ref{note:cascade}). This directly demonstrates that the readout resonator cools the first excited manifold.

Cooling is not limited to the first excited manifold. The exchange interaction also allows higher excited levels to relax through a cascade, in which successive intra-well excitations are converted into readout photons that decay at rate $\kappa_b$ (Fig.~\ref{fig:spectroscopy}a, inset). Because the resonances overlap, multiple excited levels can be cooled at a single operating point within the autoparametric band, as supported by simulations using the fitted device parameters (Supplementary Information Secs.~\ref{note:chains} and \ref{note:cascade}).

We next map the Kerr-cat bit-flip time as a function of the squeezing-drive amplitude and detuning $\Delta=\omega_a-\omega_{\mathrm{sq}}/2$ (Fig.~\ref{fig:lifetime}a). Near zero detuning, in the conventional Kerr-cat regime (Fig.~\ref{fig:lifetime}a, inset), bit-flip times remain at the millisecond scale, as in previous experiments~\cite{ding2025, adinolfi2025}. At larger detuning, however, we observe a broad region of strongly enhanced bit-flip protection, extending well beyond the autoparametric resonance associated with the readout resonator. 

Within this region, we identify a second autoparametric resonance involving the Purcell filter on the readout line, occurring when $\omega_{01}=\delta_p=\omega_{\mathrm{sq}}-\omega_p$, where $\omega_p$ is the filter frequency (purple points in Fig. 3a). Spectroscopy reveals avoided crossings between intra-well excitations and the filter mode (Supplementary Information Sec.~\ref{note:purcell}), supporting the same cooling mechanism as for the readout resonator.

At the three operating points marked in Fig.~\ref{fig:lifetime}a, we measure bit-flip times of $T_X=519\pm38$~ms on the readout resonance ($\bar n=22.4$), and $T_X=15.4\pm1.9$~s between the two resonances ($\bar n=27.3$). On the Purcell-filter resonance, reached at a higher photon number ($\bar n=33.8$), no decay is observed within the 100-s measurement window (Fig.~\ref{fig:lifetime}b). By tracking individual bit flips in repeated-readout records~\cite{berdou2023} (Supplementary Information Sec.~\ref{note:clickrecord}), we infer an idle bit-flip time of $T_X=448\pm57$~s, more than five orders of magnitude beyond previous Kerr-cat bit-flip times~\cite{adinolfi2025}.

\begin{figure}[!t]
  \centering
  \includegraphics[width=\columnwidth]{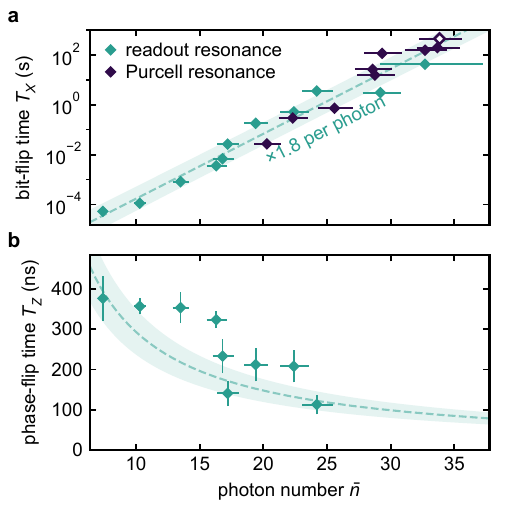}
  \caption{\textbf{Scaling bit-flip protection with photon number.}
  Measurements are performed at different flux biases, each tuned to an autoparametric cooling resonance. The mean photon number \(\bar n\) is calibrated independently by probe spectroscopy (Fig.~\ref{fig:spectroscopy}a).
  \textbf{a,} Bit-flip time $T_X$ versus $\bar n$, at the readout resonance (teal) and
  at the Purcell-filter resonance (dark purple), extracted from exponential decays of $\langle Z\rangle$ between two cat-quadrature measurements. The dashed line and the shaded region show an exponential fit and its $\pm1\sigma$
  prediction band, corresponding to a factor of $1.82\pm0.06$ increase in $T_X$ per added
  photon (Methods).
  The open diamond marks the bit-flip time $T_X=448\pm57$~s at the Purcell-filter resonance ($\bar n=33.8$), independently inferred from repeated-readout records.
  \textbf{b,} Phase-flip time $T_Z$ versus $\bar n$, measured using a Ramsey sequence. The dashed line shows the expected loss limit (Methods) for the oscillator lifetime measured at these operating points, $T_1=8.2\pm1.6$~$\mu$s; the shaded region spans the standard deviation.
  Vertical error bars denote one standard error; in \textbf{a} they are smaller than the markers. Horizontal error bars show the uncertainty of $\bar n$ from the Kerr calibration (Supplementary Information Sec.~\ref{note:nbarerror}).
  }
  \label{fig:flux}
\end{figure}

Although exponential suppression of bit flips with photon number is a defining prediction for Kerr-cat qubits, it has not previously been observed experimentally. To test whether autonomous cooling enables this scaling, we measure the bit-flip time along the autoparametric resonances at different SNAIL flux biases. Varying the flux shifts the oscillator frequency, thereby changing the minimum photon number at which each resonance can be reached (Methods). At these minimum-photon-number operating points, the bit-flip time increases approximately exponentially with $\bar n$, by a factor of $1.82\pm0.06$ per additional photon across the readout and Purcell-filter resonances (Fig.~\ref{fig:flux}a).
$T_X$ rises from $52\pm3$~$\mu$s at $\bar n=7.4$ to $448\pm57$~s at $\bar n=33.8$, an increase of approximately seven orders of magnitude, with no observed saturation.

Meanwhile, the phase-flip time decreases with photon number, remaining consistent with the single-photon-loss  limit~\cite{mirrahimi2014, rousseau2026} $T_Z \approx T_1/(2\bar n)$, with $T_1\approx8$~$\mu$s (Fig.~\ref{fig:flux}b; Methods). At the largest photon numbers, insufficient contrast in preparing coherent superpositions prevents a reliable measurement of $T_Z$ (Supplementary Information Sec.~\ref{note:phaseflipgates}). At $\bar n=24.2$, the largest photon number for which we measure both lifetimes, $T_X=3.55\pm0.23$~s and $T_Z=113\pm24$~ns, yielding a noise bias of $T_X/T_Z\approx3\times10^{7}$.

\begin{figure}[!t]
  \centering
  \includegraphics[width=\columnwidth]{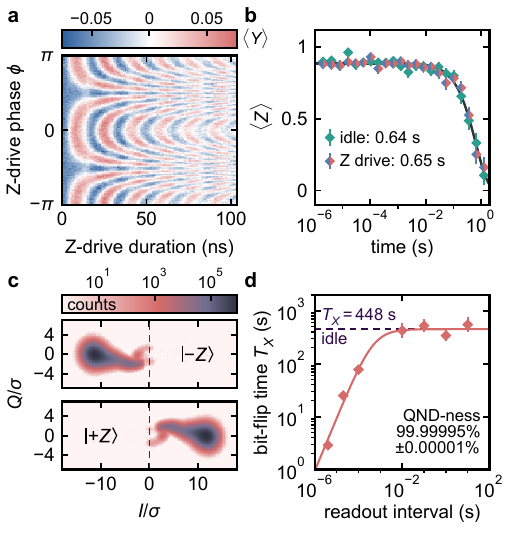}
  \caption{\textbf{Bias-preserving coherent control and readout of the autonomously cooled Kerr-cat qubit.}
  \textbf{a,} $Z$-gate Rabi oscillations at the readout-resonator autoparametric resonance ($\bar n=22.4$, teal diamond in Fig.~\ref{fig:lifetime}a). $\langle Y\rangle$ versus the duration and phase $\phi$ of a square $Z$ drive relative to the squeezing drive. The oscillations are fastest at $\phi=0$, where the drive phases are aligned, with a fitted Rabi frequency of $62.64\pm0.03$~MHz and decay time of $94\pm1$~ns.
  \textbf{b,} Bit-flip decay at the same operating point, measured without (teal) and with (blue-red diamonds) continuous $Z$ driving. Lines are exponential fits, giving $T_X=0.64\pm0.03$~s idle ($0.52$~s in Fig.~\ref{fig:lifetime}b, a separate run) and $0.65\pm0.04$~s driven. Error bars are binomial standard errors of $\langle Z\rangle$.
  \textbf{c,} Single-shot cat-quadrature readout at the Purcell-filter autoparametric resonance ($\bar n=33.8$, dark purple diamond in Fig.~\ref{fig:lifetime}a), from a
  continuous record of $2.3\times10^8$ back-to-back $4~\mu$s readouts. Readouts are grouped
  by the state heralded by the preceding readout (Methods), $|{-Z}\rangle$ (top) or
  $|{+Z}\rangle$ (bottom). Quadrature coordinates are normalised by the noise width $\sigma$, and counts are shown on a logarithmic scale. The tail between the two lobes is discussed in the Supplementary
  Information, Sec.~\ref{note:clickspectrum}.
  \textbf{d,} Bit-flip time versus the interval between repeated $4~\mu$s readouts at the same operating point. The solid curve is a fit (Methods), and the dashed line marks the idle
  bit-flip time of $448\pm57$~s. The QND-ness is one minus the fitted flip probability per readout.
  Error bars are one standard error (\textbf{b}) and 68\% Poisson intervals (\textbf{d}).}
  \label{fig:control}
\end{figure}

We next examine whether the enhanced bit-flip protection can be maintained during coherent control.
At the readout-resonance operating point (teal diamond in Fig.~\ref{fig:lifetime}a), where $T_Z=208\pm40$~ns (Fig.~\ref{fig:flux}b), a $Z$ drive~\cite{grimm2020} produces coherent oscillations in $\langle Y\rangle$ (Fig.~\ref{fig:control}a). A $Z(\pi)$ rotation takes only $8$~ns, 26 times shorter than $T_Z$. The oscillation amplitude is reduced due to dephasing during the two $204$-ns $X(\pi/2)$ gates used to prepare and measure the superposition (Supplementary Information Sec.~\ref{note:phaseflipgates}). To determine whether the $Z$ drive compromises bit-flip protection, we apply it continuously throughout the lifetime measurement. The resulting bit-flip time, $T_X=0.65\pm0.04$~s, is consistent with the idle value of $0.64\pm0.03$~s (Fig.~\ref{fig:control}b).

We also demonstrate high-fidelity cat-quadrature readout while cooling remains active. At the Purcell-filter resonance (dark purple diamond in Fig.~\ref{fig:lifetime}a), the qubit is cooled through the filter while its state is read out through the readout resonator. With a $4~\mu$s pulse, we achieve an assignment infidelity of $(5.1 \pm 0.5) \times 10^{-7}$~\cite{heinsoo2018} (Fig.~\ref{fig:control}c).
To quantify measurement-induced bit flips, we measure the bit-flip time as a function of the interval between repeated readouts (Fig.~\ref{fig:control}d). We infer a bit-flip probability of $(5\pm1)\times10^{-7}$ per measurement, corresponding to a QND-ness of $99.99995\%$~\cite{sun2014} (Methods). At intervals longer than $10$~ms, the bit-flip time remains consistent with its idle value of $448\pm57$~s.

\par\addvspace{1\baselineskip}
Multiphoton resonances between superconducting qubits and their readout circuits are
usually detrimental. They can drive qubits out of their computational
states~\cite{dumas2024, dai2026} and, in Kerr-cat qubits, sharply reduce the
protection provided by Hamiltonian confinement~\cite{benhayounekhadraoui2026}. Here we have shown that, when the readout modes are sufficiently cold and their frequencies lie below that of the squeezing drive, these resonances can instead provide autonomous cooling. 
The bit-flip time increases exponentially with photon number and reaches 450~s, bringing Kerr-cat protection into the regime previously achieved only through dissipative stabilisation. At operating points with bit-flip times up to a few seconds, phase-flip rates remain consistent with intrinsic single-photon loss, with a measured noise bias reaching $T_X/T_Z\approx3\times10^{7}$
(Fig.~\ref{fig:flux}).
We also demonstrate fast coherent control of the cooled qubit, performing $Z$ gates on a 10-ns timescale without measurable degradation of the bit-flip lifetime. Repeated readout introduces a bit-flip probability of $(5\pm1)\times 10^{-7}$ per measurement.

With bit-flip rates now strongly suppressed, reducing phase-flip errors becomes increasingly important. The phase-flip time is limited by photon loss to approximately $T_Z=T_1/(2\bar{n})$ (Methods), making the oscillator's single-photon lifetime particularly important at the large photon numbers used here. Our SNAIL oscillator has a single-photon lifetime of approximately $8$~$\mu$s, whereas lifetimes of $55.7$~$\mu$s have already been demonstrated~\cite{adinolfi2025}. This suggests ample room to improve phase coherence through device design and fabrication.
Colder circuitry could provide a complementary improvement by reducing the residual excited-state population, potentially allowing comparable bit-flip protection at smaller photon numbers (Supplementary Information Sec.~\ref{note:floor}).

By overcoming the longstanding bit-flip limitation of Kerr-cat qubits without additional cooling drives or circuit elements, autonomous cooling brings strong intrinsic protection together with fast coherent control. These results make Kerr-cat qubits a promising platform for quantum error correction that exploits their strongly biased noise.

\bibliography{references_short}

\clearpage

\section*{Methods}

\subsection*{Device and experimental setup}

The device was fabricated by angle evaporation of aluminium onto a 381-µm-thick intrinsic silicon wafer (Siegert Wafer). The device consists of a flux-tunable SNAIL
oscillator~\cite{frattini2017} with frequency $\omega_a/2\pi = 3.667$ to $3.789$~GHz over the
flux biases investigated, coupled to a readout resonator
(frequency $\omega_b/2\pi = 7.297$~GHz, linewidth $\kappa_b/2\pi = 0.645$~MHz). The readout resonator is coupled to the output line through a Purcell
filter (frequency $\omega_p/2\pi = 7.187$~GHz, linewidth $\kappa_p/2\pi = 0.85$~MHz). The readout resonator
lies below twice the SNAIL frequency, with a detuning $L = 2\omega_a - \omega_b$. A microwave drive at frequency $\omega_{\mathrm{sq}}$ is applied through the readout port, displacing the readout resonator. Through the three-wave exchange interaction $g_{21}$, this field generates the two-photon squeezing drive $\varepsilon_2$ on the SNAIL oscillator (Eq.~\eqref{eq:kcq}). A magnetic flux
hose~\cite{navau2014, sanchez2021} positioned above the chip threads a magnetic flux $\Phi_{\mathrm{ext}}$
through the SNAIL loops, tuning $\omega_a$ and hence $L$. Across the flux biases used in
Fig.~\ref{fig:flux}, $L/2\pi$ ranges from $36.2$ to $281.5$~MHz, while the Kerr nonlinearity $K/2\pi$ varies from $1.52$ to
$1.99$~MHz (Supplementary Information Fig.~\ref{sfig:snailflux} and
Table~\ref{stab:parameters}). At the flux bias used in Fig.~\ref{fig:lifetime}, the three-wave exchange coupling between the SNAIL oscillator and the readout resonator is
$g_{21}/2\pi = 0.74 \pm 0.1$~MHz. Further details of device fabrication, packaging, and wiring are provided in Supplementary Information Sec.~\ref{note:device}.

\subsection*{Hamiltonian and spectroscopy model}

We describe the SNAIL oscillator and readout resonator, with annihilation operators $\hat a$ and $\hat{\tilde b}$, respectively, using a two-mode Hamiltonian derived from the circuit Hamiltonian in Supplementary Information
Sec.~\ref{note:hamiltonian}. In a frame rotating at $\omega_{\mathrm{sq}}/2$ for the
SNAIL oscillator and at $\omega_{\mathrm{sq}}$ for the readout resonator, the Hamiltonian is
\begin{align}
  \hat H
  &= \Delta\,\hat a^\dagger\hat a - K\,\hat a^{\dagger 2}\hat a^{2}
     - \delta_b\,\hat{\tilde b}^\dagger\hat{\tilde b}
     + g_{21}\big(\hat a^{2}\hat{\tilde b}^\dagger + \hat a^{\dagger 2}\hat{\tilde b}\big) \nonumber\\
  &\quad - \chi_{ab}\,\hat a^\dagger\hat a\,\hat{\tilde b}^\dagger\hat{\tilde b}
     + \varepsilon_{\mathrm{sq}}\big(\hat{\tilde b} + \hat{\tilde b}^\dagger\big),
  \label{eq:twomode}
\end{align}
where $\Delta = \omega_a - \omega_{\mathrm{sq}}/2$ and
$\delta_b = \omega_{\mathrm{sq}} - \omega_b$ are the SNAIL and readout detunings, respectively, $\chi_{ab}$ is the
cross-Kerr coupling, and $\varepsilon_{\mathrm{sq}}$ is the squeezing-drive amplitude applied to the readout resonator. The readout resonator loses photons at rate $\kappa_b$.

The squeezing drive produces a steady-state coherent field $\beta = \varepsilon_{\mathrm{sq}}/(\delta_b + i\kappa_b/2)$ in the readout resonator. We therefore write $\hat{\tilde b} = \beta + \hat b$, where $\hat b$ describes fluctuations around this field. The coherent field generates an effective two-photon
squeezing drive $\varepsilon_2=g_{21}\beta$ on the
SNAIL oscillator through the three-wave exchange
interaction, while the cross-Kerr interaction produces
a Stark shift.  Collecting the terms independent of
$\hat b$ gives the effective Kerr-cat Hamiltonian
\begin{align}
  \hat H_{\mathrm{KCQ}}
  &= \big(\Delta - \chi_{ab}|\beta|^2\big)\hat a^\dagger\hat a
     - K\,\hat a^{\dagger 2}\hat a^{2} \nonumber\\
  &\quad + \varepsilon_2^*\hat a^{2} + \varepsilon_2\hat a^{\dagger 2},
  \label{eq:kcq}
\end{align}
which produces the double-well potential that confines the cat qubit~\cite{puri2017, grimm2020}.
Choosing the phases of $\hat a$ and $\hat b$ makes $\varepsilon_2$ and $\beta$ real, as we take them from here on.
The remaining terms, containing the displaced readout operator
$\hat b$, are given in Supplementary Information
Eq.~\eqref{eq:twomodedisplaced}.

To model the excited-state spectroscopy (dashed lines in Fig.~\ref{fig:spectroscopy}a) and
calibrate the mean photon number $\bar n$ (Figs.~\ref{fig:lifetime}a
and~\ref{fig:flux}), we numerically diagonalise the displaced two-mode Hamiltonian derived
from Eq.~\eqref{eq:twomode} as a function of the drive amplitude $\varepsilon_{\mathrm{sq}}$.
We fit the measured transition frequencies to the corresponding eigenenergy differences, obtaining a calibration between the applied room-temperature drive amplitude and the dimensionless squeezing strength
$\varepsilon_2/K$. The calibrated Hamiltonian is then used to determine the mean photon number $\bar n$ of the ground-state cat manifold (Supplementary Information Sec.~\ref{note:calibration}).

\subsection*{Autoparametric exchange interaction}

The exchange interaction between the SNAIL oscillator
and the displaced readout mode,
$g_{21}(\hat a^2\hat b^\dagger+\mathrm{h.c.})$,
takes a particularly simple form in the shifted Fock
basis~\cite{xu2022,putterman2022}. Writing $\hat a\approx\hat Z\otimes(\alpha+\hat d)$,
where $|\alpha|^2=\bar n$, $\hat d$ is the intra-well
annihilation operator, and $\hat Z$ acts on the logical qubit, the exchange interaction becomes
\begin{align}
  g_{21}\big(\hat a^{2}\hat b^\dagger + \hat a^{\dagger 2}\hat b\big)
  &\approx \hat I\otimes\Big[g_{21}\bar n\big(\hat b + \hat b^\dagger\big) \nonumber\\
  &\qquad + 2g_{21}\sqrt{\bar n}\big(\hat d\,\hat b^\dagger + \hat d^\dagger\hat b\big) \nonumber\\
  &\qquad + g_{21}\big(\hat d^{2}\hat b^\dagger + \hat d^{\dagger 2}\hat b\big)\Big].
  \label{eq:vertices}
\end{align}
The three terms describe processes involving zero,
one, or two intra-well excitations and one readout
photon. Within this approximation, all three act
as the identity on the logical qubit. The
squeezing-drive frequency, through the detuning $\delta_b$, determines which process is resonant. The first term drives the readout mode without changing the oscillator state. The
second term describes the beam-splitter interaction between the intra-well and readout modes (Eq.~\eqref{eq:beamsplitter}), which becomes resonant at $\omega_{01} = \delta_b$. The third term converts two intra-well excitations into
one readout photon and becomes resonant when $\omega_{02} = \delta_b$, with $\omega_{02}$ the
energy separation between the ground and second excited doublets. The two-excitation resonance \(\omega_{02}=\delta_b\) occurs at a lower photon number than the one-excitation resonance \(\omega_{01}=\delta_b\).
The displaced cross-Kerr interaction modifies the
zero- and one-excitation coupling amplitudes by
relative corrections of order
$\chi_{ab}|\beta|/g_{21}$, but leaves the two-excitation term unchanged.

A similar analysis applies to the Purcell filter. Because the filter mode hybridises with the readout resonator, the SNAIL's
cubic nonlinearity also mediates a two-to-one exchange between the SNAIL oscillator and the filter. In the shifted Fock basis, this interaction has the same form as Eq.~\eqref{eq:vertices}, but with a different effective coupling strength. It provides an additional autonomous cooling channel when the first intra-well transition is resonant with the filter, $\omega_{01}=\delta_p$, where $\delta_p=\omega_{\mathrm{sq}}-\omega_p$ (Figs.~\ref{fig:lifetime}a,b).

\subsection*{Dynamical measurement and readout}

Except for the excited-state relaxation measurements in Fig.~\ref{fig:spectroscopy}d, which use dispersive readout, all measurements
distinguish the left and right well states,
$|\psi_0^L\rangle$ and $|\psi_0^R\rangle$, using
cat-quadrature readout~\cite{grimm2020}. With the squeezing drive kept on, we apply a
$4~\mu$s conversion drive at $\omega_b - \omega_a$, activating a beam-splitter interaction
that displaces the readout resonator with opposite phases for the two wells. The states are distinguished in a single shot by demodulating the transmitted readout field.

To maintain a single-shot assignment fidelity above $99.9\%$ throughout the investigated parameter space,
including regions where the readout is degraded by
autoparametric hybridisation, we use a dynamical
measurement protocol that separates the readout and
operating points. The readout point has the same
squeezing-readout detuning $\delta_b$ as the operating
point, but a higher squeezing amplitude, at which the
double-well barrier is deeper and the readout fidelity
is higher. Following the first readout, we adiabatically ramp the squeezing amplitude to the operating point over  $1.2~\mu$s, hold it for a variable duration,
and ramp back over $1.2~\mu$s before the second readout. From repeated sequences, we
extract $\langle Z_0Z_1\rangle$, the mean product of the two readout outcomes, not corrected for readout infidelity
(Fig.~\ref{fig:spectroscopy}a). This product is the expectation value $\langle Z\rangle$, with $+Z$ defined as the well heralded by the first readout (Fig.~\ref{fig:lifetime}b). For the repeated-readout measurements in Figs.~\ref{fig:control}c,d, the readout is performed directly at the operating point, without ramps.

\subsection*{Excited-state spectroscopy and decay}

For the two-tone spectroscopy in Fig.~\ref{fig:spectroscopy}a–c, we initialise the qubit in
one well using a cat-quadrature readout, adiabatically ramp the squeezing amplitude to the target
value over $1.2~\mu$s, and apply a weak probe tone at
a variable frequency near $\omega_{\mathrm{sq}}/2$.
We then ramp back over $1.2$~$\mu$s and perform a
second cat-quadrature readout. When the probe is resonant with an intra-well transition, it
populates excited states with higher interwell switching rates, reducing $\langle Z_0Z_1\rangle$. To resolve transitions to higher excited levels, we gradually increase the probe amplitude at larger
negative probe detunings.

To measure the relaxation time $T_1$ of the first excited manifold across the level anticrossing
(Fig.~\ref{fig:spectroscopy}d,e), we use dispersive excited-state readout~\cite{adinolfi2025}.
We first ramp the squeezing amplitude over $1.2$~$\mu$s to a preparation point on either
side of the autoparametric band and allow the oscillator to
relax into the ground manifold.
A resonant $\pi$ pulse then prepares the first excited
manifold $|\psi_1^\pm\rangle$.
We ramp the squeezing amplitude to its target value,
hold it for a variable duration, and ramp
back to the preparation point.
These ramps each take 120 ns and are approximately adiabatic.
After a $200$-ns settling time, we measure the
remaining excited-state population dispersively.
Each relaxation trace is fitted to a
single exponential decay to extract $T_1$ and its standard error.

\subsection*{Logical lifetime measurements}

To measure the bit-flip time $T_X$ (Figs.~\ref{fig:lifetime}a,b and~\ref{fig:flux}a), we use
the dynamical measurement sequence presented above without any probe tone, varying the hold time $t$ at the target operating point before ramping back to the second cat-quadrature
readout. Each trace $\langle Z(t)\rangle$ is fitted to a single exponential
decay $A\exp(-t/T_X)$ with zero offset, reflecting the symmetry between the two wells at
$t \to \infty$.

To measure the phase-flip time $T_Z$ across photon numbers (Fig.~\ref{fig:flux}b), we use
a Ramsey sequence that heralds a well state with a first cat-quadrature readout and applies
a Kerr refocusing $X(\pi/2)$ gate to prepare an equal superposition of
the two wells, followed after a hold time $t$ by a second $X(\pi/2)$
gate and a second readout (Supplementary Information Sec.~\ref{note:phaseflipgates}). This
gate~\cite{kirchmair2013, grimm2020} switches off the two-photon drive for $\pi/(2K)$ so that the state evolves under the Kerr nonlinearity alone. Because
$T_Z$ at large $\bar n$ can be shorter than $\pi/(2K)$, performing the gate at
the large-$\bar n$ operating point would cause severe dephasing.
We therefore change the squeezing amplitude dynamically during the sequence, applying the gates at a lower amplitude and hence a lower photon number.
Between the gates, we ramp the amplitude to the operating point for the hold time $t$ and back, with each ramp taking $48$ to $64$~ns.
The resulting decay of $\langle Y\rangle$ is fitted to a single exponential to extract $T_Z$.

We cross-check this dynamical protocol against a phase-modulated $X(\pi/2)$ gate, a modified version of a previously demonstrated detuning gate~\cite{hajr2024}. The gate lowers the double-well barrier transiently by modulating only the phase of the two-photon drive, without switching off the squeezing drive or ramping its amplitude (Supplementary Information
Sec.~\ref{note:phaseflipgates} and Fig.~\ref{sfig:phaseflipgates}).
At the $\bar n = 13.5$ autoparametric resonance of Fig.~\ref{fig:flux}b, the phase-modulated gate, applied directly at the operating point, gives $T_Z = 351 \pm 67$~ns, and the dynamical Kerr refocusing gate gives $354 \pm 45$~ns.

We compare the measured $T_Z$ with the limit set by single-photon loss and gain~\cite{mirrahimi2014, rousseau2026}, both of which flip the parity of the cat:
\begin{equation}
  T_Z = \frac{T_1}{2[(1+2n_{\mathrm{th},a})\bar n + n_{\mathrm{th},a}]},
  \label{eq:tzlimit}
\end{equation}
where $n_{\mathrm{th},a}$ is the oscillator's thermal occupation. For $n_{\mathrm{th},a} = 0$ this becomes the simplified formula $T_1/(2\bar n)$ used above.
The dashed line in Fig.~\ref{fig:flux}b evaluates Eq.~\eqref{eq:tzlimit} at the occupation $\bar n_{\mathrm{th}} = 0.185$ inferred from the excited-state population, which also includes quantum heating and is therefore an upper limit on $n_{\mathrm{th},a}$ (Supplementary Information Sec.~\ref{note:thermometry}).
The line is therefore a lower limit on the loss-limited $T_Z$.

\subsection*{Scaling of logical errors with photon number}

At a fixed flux bias, an autoparametric resonance with the readout resonator or the
Purcell filter can be reached only above a minimum number of photons. For the readout resonance, varying the squeezing frequency changes both $\delta_b$ and $\Delta$, while their combination $\delta_b+2\Delta=L$ remains fixed. Together with the resonance condition $\omega_{01}=\delta_b$, this imposes a minimum photon number of approximately $L/(5K)$ (Supplementary Information Sec.~\ref{note:gap}). The same holds
for the Purcell-filter resonance, with $\omega_p$ in place of $\omega_b$. To reach the
resonances at lower photon numbers, we changed the flux bias of the SNAIL, which tunes
$\omega_a$ and with it $L$. Each point in Fig.~\ref{fig:flux} was taken at a different
flux bias, at the minimum photon number of the readout or the Purcell-filter resonance.
At each point, we calibrated $\bar n$ from probe spectroscopy with the model of
Eq.~\eqref{eq:twomode} and measured $T_X$ and $T_Z$ at the same squeezing frequency and
amplitude, as described above.

To quantify how fast $T_X$ grows with $\bar n$, we fitted a straight line to $\ln T_X$
against $\bar n$ by unweighted least squares, over the points of both resonances in
Fig.~\ref{fig:flux}a.
The fitted slope is $0.60\pm0.03$ (one standard error), a factor of $1.82\pm0.06$ per photon. We estimated the standard error from the scatter of the points about the line, which reflects the errors in $\bar n$ and $T_X$ at each flux bias as well as variation between operating points.
The shaded region in Fig.~\ref{fig:flux}a is the $\pm1\sigma$ prediction band of this fit: a new measurement at a given $\bar n$ would fall inside it with $68\%$ probability.

Changing the flux bias also changes other parameters of the oscillator. Across the flux biases of Fig.~\ref{fig:flux}, $K$ and the single-photon lifetime $T_1$ vary, while $g_{21}$ changes with the cubic nonlinearity of the SNAIL (Supplementary Information Table~\ref{stab:parameters}). Although these variations can affect the absolute bit-flip rates, they do not obscure the observed exponential increase: the points of both resonances approximately follow a common exponential dependence on $\bar n$ over nearly seven orders of magnitude in $T_X$ (Fig.~\ref{fig:flux}a).

\subsection*{Bias-preserving $Z$ drive}

The $Z$ drive of Fig.~\ref{fig:control} is a square pulse at $\omega_{\mathrm{sq}}/2$,
delivered through the pin under the SNAIL resonator
(Supplementary Information Sec.~\ref{note:device}), with its phase $\phi$ set relative to
the squeezing drive up to a calibrated offset.
For the map of Fig.~\ref{fig:control}a, we used the dynamical measurement sequence at the
readout-resonance operating point.
Between the two cat-quadrature readouts, a phase-modulated $X(\pi/2)$ gate (Supplementary
Information Sec.~\ref{note:chirpgate}) prepared a cat state, the $Z$ drive acted for a
duration $\Delta t$, and a second phase-modulated $X(\pi/2)$ gate mapped the state back onto the
wells.
At $\Delta t = 0$, the two gates compose an $X(\pi)$ gate, which switches the well.
We fitted the whole map with a damped cosine whose frequency scales as $\cos\phi$, which
gave the Rabi frequency at the fastest phase and the decay time quoted in
Fig.~\ref{fig:control}a.

For the bit-flip decays of Fig.~\ref{fig:control}b, we used the $T_X$ sequence of the logical
lifetime measurements at the same operating point and measured every hold time twice, once
idle and once with the $Z$ drive on for the whole hold.
We interleaved the two measurements so that slow fluctuations of the bit-flip time affected
both equally.
The $Z$ drive was applied at the amplitude and phase of the fastest oscillation in
Fig.~\ref{fig:control}a, where a $Z(\pi)$ rotation takes $8$~ns.
We fitted the idle and the driven decays each to a single exponential with zero offset.

\subsection*{Measurement-induced bit flips}

To quantify measurement-induced bit flips (Fig.~\ref{fig:control}d), we recorded
continuous sequences of $4~\mu$s cat-quadrature readouts, repeated at an interval $D$ from
$4~\mu$s (back-to-back) to $10$~s. We acquired these records at the operating point
marked by the dark purple diamond in Fig.~\ref{fig:lifetime}a and the open diamond in Fig.~\ref{fig:flux}a, on the
autoparametric resonance with the Purcell filter ($\bar n = 33.8$). At this operating point, the readout resonator is sufficiently detuned from the intra-well transitions that the dynamical measurement protocol described above is unnecessary. We therefore performed readout directly at the operating point, without ramps, keeping the qubit at the same operating point throughout each record (Supplementary Information
Sec.~\ref{note:clickrecord}). In total, we counted $62$ flips between the two wells in
$15$~h for $D>10~$ms, where readout-induced bit-flips are negligible. For each readout interval, we calculate $T_X$ as the total exposure time divided by twice the number of observed flips. Each readout
flips the qubit with a probability $p_{\mathrm{flip}}$ in addition to the idle flips, so
that $1/T_X = 1/T_X^0 + 2p_{\mathrm{flip}}/D$, with $T_X^0$ the idle bit-flip
time~\cite{sun2014}. A Poisson maximum-likelihood fit to the flip counts of all readout intervals
gives $T_X^0 = 448\pm57$~s and $p_{\mathrm{flip}} = (5.2\pm0.3)\times10^{-7}$. Estimating $p_{\mathrm{flip}}$ separately for each readout interval $D \le 100~\mu$s gives values ranging from $3.8\times10^{-7}$
to $6.9\times10^{-7}$, a scatter larger than expected from Poisson noise (deviance of $25.5$ for $5$ degrees of freedom). To account for this excess scatter, we scale the uncertainty of
$p_{\mathrm{flip}}$ by $\sqrt{25.5/5}$ and quote $p_{\mathrm{flip}} = (5\pm1)\times10^{-7}$, corresponding to a QND-ness of $1 - p_{\mathrm{flip}}=99.99995\%$.

\subsection*{Readout assignment infidelity}

We measured the assignment infidelity of Fig.~\ref{fig:control}c on the back-to-back ($D=4~\mu$s) record
of the measurement-induced bit-flip experiment, with the readouts themselves serving as heralds.
Each readout gives a point $(I,Q)$ in the plane of Fig.~\ref{fig:control}c, with $I$ along
the axis joining the two readout lobes, whose centers lie at $I=\pm h$, approximately $\pm11\sigma$
(Supplementary Information Sec.~\ref{note:clickrecord}).
Here $\sigma$ is the standard deviation of a Gaussian fitted to the lobe core along $I$.
A readout heralded $|{-Z}\rangle$ when $I<-h/2$ and $|{+Z}\rangle$ when $I>h/2$.
Of all readouts, $97.6\%$ heralded a state, while the remaining $2.4\%$ were not used as heralds.
We assigned the readout following a herald to $|{-Z}\rangle$ or $|{+Z}\rangle$ by the sign of $I$.
The assignment infidelity is the mean of the two error probabilities~\cite{heinsoo2018},
$[p(+Z|{-Z})+p(-Z|{+Z})]/2$, where $p(+Z|{-Z})$ is the probability that a $|{-Z}\rangle$
herald is followed by a $|{+Z}\rangle$ assignment.
We took its uncertainty from the Poisson statistics of the error counts.

\subsection*{Use of large language models (LLMs)}

LLMs were used throughout this work. They assisted in drafting and editing the
manuscript and the Supplementary Information, and in writing and debugging the code for
the experiment and the data analysis. The authors reviewed, verified, and revised all
text, code, and results produced with their assistance.

\section*{Data availability}
The data that support the findings of this study are available from the corresponding authors upon reasonable request.

\section*{Code availability}
The code used in this study is available from the corresponding authors upon reasonable request.

\section*{Acknowledgements}
We thank R. Leibovitz for numerical simulations and for contributing to the development
of the analytical model. We thank G. Kirchmair and P. Straub for discussions on the
design of the magnetic flux hose.
This work was supported by QuamCore Ltd. through a sponsored research agreement with Yeda Research and Development Co. Ltd. (agreement no. 11355). S.R. is the incumbent of the Rabbi Dr. Roger Herst Career Development Chair.

\section*{Author contributions}
S.H. and M.S. contributed equally to this work. S.H., M.S., F.L. and S.R. conceived the project. S.H. and M.S. designed and performed the experiments and analysed the data. M.S. fabricated the devices. U.G. and B.G. contributed to the development of software infrastructure and calibration tools. S.H., M.S., F.L. and S.R. wrote the manuscript, with input from all authors. S.R. and F.L. supervised the project.

\section*{Competing interests}
S.R. is a shareholder of QuamCore Ltd., which funded part of this work. The remaining authors declare no competing interests. 
\clearpage

\onecolumngrid

\setcounter{section}{0}
\setcounter{figure}{0}
\setcounter{table}{0}
\setcounter{equation}{0}
\renewcommand{\thesection}{S\arabic{section}}
\renewcommand{\thefigure}{S\arabic{figure}}
\renewcommand{\thetable}{S\arabic{table}}
\renewcommand{\theequation}{S\arabic{equation}}

\begin{center}
  \large\bfseries
  Supplementary Information:\\
  Autonomous cooling of Kerr-cat qubits to minutes-long bit-flip times
\end{center}

\makeatletter
\let\l@section@orig\l@section
\let\l@subsection@orig\l@subsection
\newcommand{\suppltochide}{\let\l@section\@gobble@tw@\let\l@subsection\@gobble@tw@}
\newcommand{\suppltocshow}{\let\l@section\l@section@orig\let\l@subsection\l@subsection@orig}
\makeatother
\tableofcontents
\addtocontents{toc}{\protect\suppltocshow}
\vspace{1em}

\let\supplsection\section
\renewcommand{\section}[1]{\supplsection{#1}\suppressfloats[t]}
\renewcommand{\bottomfraction}{0.7}

\section{Device and experimental setup}
\label{note:device}

\begin{figure}[tbp]
  \centering
  \includegraphics[width=0.7\textwidth]{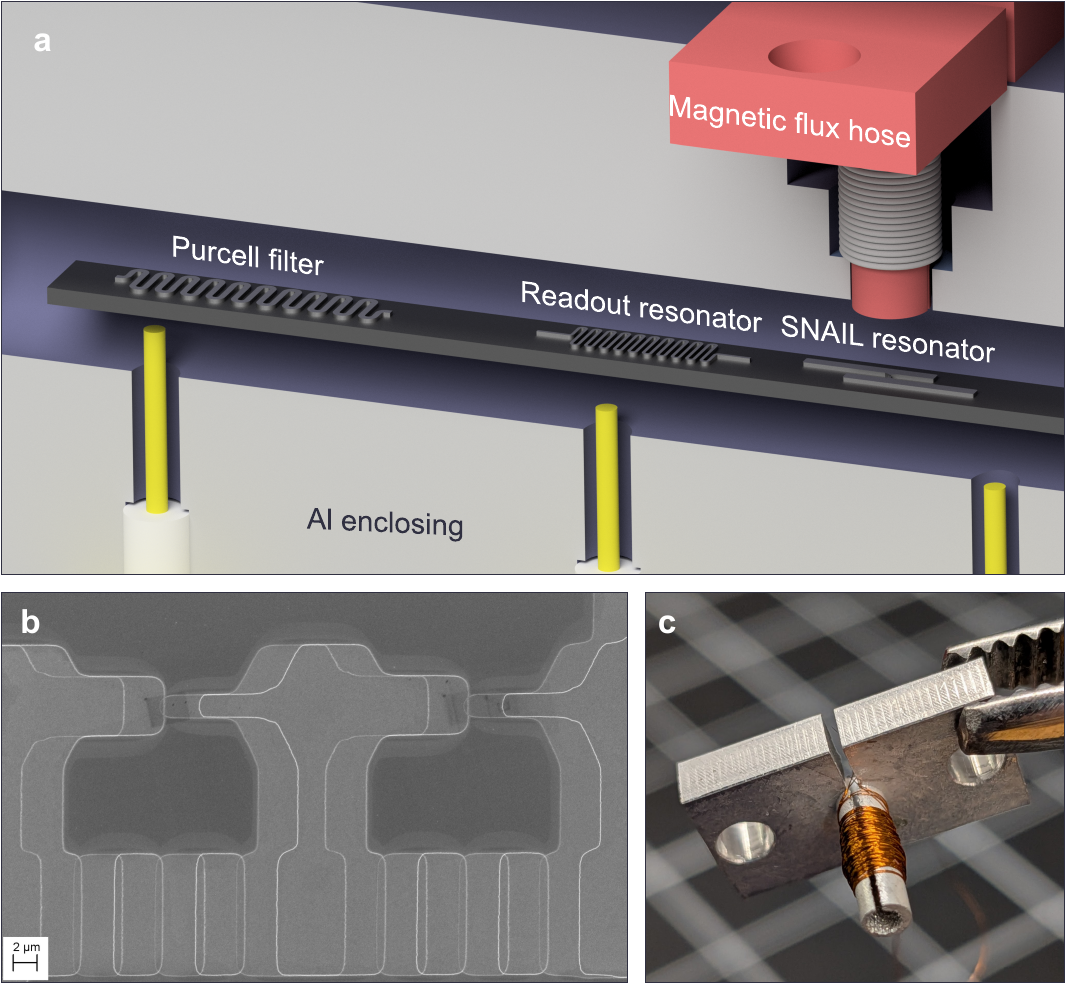}
  \caption{\textbf{Device and packaging.} \textbf{a,} Rendering of the chip inside its aluminium
  enclosure. From left to right, the chip carries a Purcell filter, the readout resonator
  (mode $\hat b$), and the SNAIL oscillator (mode $\hat a$); the device parameters
  are listed in Table~\ref{stab:parameters}. Coaxial pins coupling to the chip from below are
  drawn in yellow: the pin under the Purcell filter serves the readout input and output lines,
  the pin under the readout resonator delivers the squeezing drive $\varepsilon_{\mathrm{sq}}$,
  and the pin under the SNAIL resonator carries the combined drive line ($\varepsilon_{\mathrm{probe}}$,
  $\varepsilon_{\mathrm{CQR}}$, and $\varepsilon_x$). A magnetic flux hose mounted above the chip
  delivers the DC flux $\Phi_{\mathrm{ext}}$ threading the SNAIL loops, which sets $\omega_a$ and
  with it the detuning $L = 2\omega_a - \omega_b$ that controls the cooling resonance across
  Fig.~\ref{fig:flux} of the main text. \textbf{b,} Scanning electron micrograph of a double SNAIL element, formed by an
  array of three large Josephson junctions shunted by a single smaller junction in a superconducting
  loop. \textbf{c,} Photograph of the magnetic flux hose assembly with the flux hose holder surrounded by a spool of NbTi wire inside a copper matrix}
  \label{sfig:device}
\end{figure}

In this section we describe the sample, its fabrication, and the measurement chain used
throughout the experiment. The device and its packaging are shown in
Fig.~\ref{sfig:device}, and all measured, design, and derived system parameters are
summarised in Table~\ref{stab:parameters}.

\begin{table}[!h]
  \centering

\caption{\textbf{Measured, designed, and derived device parameters.} Flux-dependent parameters are reported at the nominal operating point ($5.70$\,V) and across the full flux window ($5.55$ to $6.10$\,V; Fig.~\ref{fig:flux} of the main text). Corresponding main-text figures for each flux bias are listed in the figure column, and the oscillator frequency is also quoted over its full tuning span. For $T_1$, $\pm$ represents the standard deviation of calibrations at $5.70$\,V, while the quoted range spans the per-flux means shown in Fig.~\ref{sfig:snailflux}c. The uncertainty in $g_{21}$ indicates its variation across spectroscopy maps. The lower section details the SNAIL circuit: design values alongside a circuit-model fit to the flux-dependent oscillator frequency and Kerr coefficient (Fig.~\ref{sfig:snailflux}). In this fit, the junction ratio, capacitance, and flux calibration were left unconstrained while inductances were held at design values; $g_3$ is extracted from the same fit. All fit uncertainties represent one standard error.}

  \label{stab:parameters}
  \newcommand{\figop}{\ref{fig:spectroscopy}, \ref{fig:lifetime}, \ref{fig:control}}
  \newcommand{\figwin}{\ref{fig:flux}}
  \begin{ruledtabular}
  \begin{tabular}{llllll}
    Symbol & Description & Value & Obtained from & Flux bias & Main fig. \\
    \hline
    $\omega_a/2\pi$      & oscillator frequency             & $3.757$\,GHz & Two-tone spectroscopy & $5.70$\,V & \figop \\
                         &                                  & $3.667$–$3.789$\,GHz & & $5.55$–$6.10$\,V & \figwin \\
                         &                                  & $3.41$–$4.61$\,GHz & & full tuning & — \\
    $\omega_b/2\pi$      & readout frequency                & $7.298$\,GHz & Single-tone spectroscopy & $5.70$\,V & \figop \\
                         &                                  & $7.297$–$7.298$\,GHz & & $5.55$–$6.10$\,V & \figwin \\
    $\omega_p/2\pi$      & Purcell filter frequency         & $7.187$\,GHz & Single-tone spectroscopy & & \\
    $L/2\pi$             & mode detuning, $2\omega_a - \omega_b$ & $216$\,MHz & $\omega_a, \omega_b$ spectroscopy & $5.70$\,V & \figop \\
                         &                                  & $36.2$–$281.5$\,MHz & & $5.55$–$6.10$\,V & \figwin \\
    $K/2\pi$             & oscillator Kerr coefficient      & $1.97$\,MHz & Oscillator anharmonicity & $5.70$\,V & \figop \\
                         &                                  & $1.52$–$1.99$\,MHz & & $5.55$–$6.10$\,V & \figwin \\
    $g_3/2\pi$           & bare cubic coefficient           & $10.8 \pm 0.9$\,MHz & SNAIL circuit fit & $5.70$\,V & \figop \\
                         &                                  & $10.6$–$11.3$\,MHz & & $5.55$–$6.10$\,V & \figwin \\
    $g_{21}/2\pi$        & two-to-one exchange rate         & $0.74 \pm 0.10$\,MHz & Two-mode fit (Sec.~\ref{note:twomodefit}) & $5.70$\,V & \figop \\
                         &                                  & $0.60$–$0.85$\,MHz & & $5.55$–$6.10$\,V & \figwin \\
    $\kappa_b/2\pi$      & readout decay rate               & $0.645$\,MHz & Readout linewidth fit & & \\
    $\kappa_p/2\pi$      & Purcell filter decay rate        & $0.85$\,MHz & Purcell linewidth fit & & \\
    $T_1$                & oscillator single-photon lifetime & $6.9 \pm 0.8\,\mu$s & Free decay & $5.70$\,V & \figop \\
                         &                                  & $3.4$–$9.9\,\mu$s & & $5.55$–$6.10$\,V & — \\
    $p_1$                & population of $|\psi_1^\pm\rangle$, driven & $12.6$–$13.9\%$ & Thermometry (Sec.~\ref{note:thermometry}) & & \\
    \hline
    —                    & SNAILs in series                 & two & Design & & \\
    $L_{\mathrm{J}}$     & large-junction inductance        & $1.11$\,nH & Design & & \\
    $L_{\mathrm{lin}}$   & series linear inductance         & $0.88$\,nH & Design & & \\
    $\alpha_{\mathrm{J}}$ & small-to-large junction ratio   & $0.12$ & Room-temperature resistance & & \\
                         &                                  & $0.112$ & Circuit fit & & \\
    $E_C/h$              & oscillator charging energy       & $87.3$\,MHz & Q3D simulation & & \\
                         &                                  & $82$\,MHz & Circuit fit & & \\
  \end{tabular}
  \end{ruledtabular}
\end{table}

The device is composed of a single $33.45 \times 1.8$\,mm chip held in a $4$-mm-diameter tunnel of circular cross section
machined into a 6061 aluminium enclosure, Fig.~\ref{sfig:device}a. The
package was mounted on the mixing chamber plate of a dilution refrigerator operating at $10$~mK.

The circuit is fabricated in a single step on a $381~\mu$m-thick intrinsic silicon substrate (Siegert Wafer)
by electron-beam lithography operating at $150$\,nA using a trilayer resist stack (LOR~20C undercut layer,
PMMA~950K~A5 imaging layer, and an AR-PC~5090 / Electra~92 conductive top coat), followed by
double-angle shadow evaporation of aluminium inside a Plassys electron-beam evaporator. The chip
carries three elements, ordered from left to right in Fig.~\ref{sfig:device}a: a Purcell filter,
a readout resonator, and a SNAIL resonator incorporating two SNAIL elements
(Fig.~\ref{sfig:device}b). The Purcell filter sits between the readout resonator and the
readout line, passing the readout signal while suppressing the SNAIL oscillator's decay into
that line.

Three coaxial pins couple capacitively to the chip from below, drawn in yellow in
Fig.~\ref{sfig:device}a. From left to right, the pin under the Purcell filter serves the
readout input and output; the pin under the readout resonator carries the squeezing drive
$\varepsilon_{\mathrm{sq}}$, which drives the SNAIL resonator indirectly through the readout mode;
and the pin under the SNAIL resonator delivers the single-photon control, spectroscopy, and
conversion tones ($\varepsilon_{\mathrm{probe}}$, $\varepsilon_{\mathrm{CQR}}$, and $\varepsilon_x$).

\begin{figure}[tbp]
  \centering
  \includegraphics[width=\textwidth]{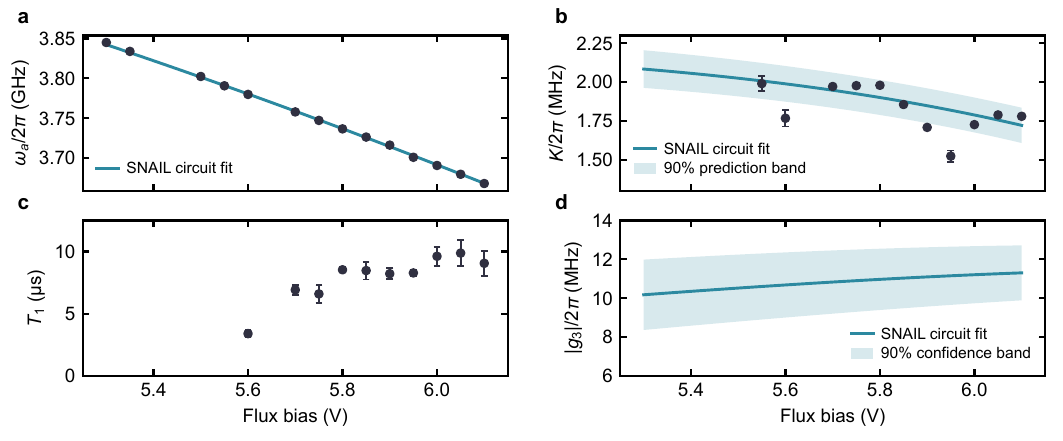}
  \caption{\textbf{SNAIL oscillator against flux bias.} \textbf{a,} Oscillator frequency $\omega_a$.
  \textbf{b,} Kerr coefficient $K$. \textbf{c,} Single-photon lifetime $T_1$. \textbf{d,} Bare
  cubic coefficient $|g_3|$, inferred from the fit. The lines in \textbf{a}, \textbf{b} and
  \textbf{d} are one fit of the SNAIL circuit model to $\omega_a$ and $K$ together, with the
  junction ratio, the capacitance and the flux calibration free and the inductances at their
  design values (Table~\ref{stab:parameters}). The points of \textbf{b} at $5.60$ and
  $5.95$\,V are outliers and are not included in the fit. The band in \textbf{b} is the $90\%$ prediction
  band for a single calibration of $K$, and the band in \textbf{d} the $90\%$ confidence band of
  the fit. Points are the mean over the calibrations at each flux bias, with error bars of one
  standard error, or of the fit where a flux bias has a single value.}
  \label{sfig:snailflux}
\end{figure}

Sweeping the detuning $L = 2\omega_a - \omega_b$ requires a static flux $\Phi_{\mathrm{ext}}$
through the SNAIL loops, which is more challenging to deliver in a three-dimensional superconducting architecture
than in a planar circuit, where a flux line sits on the chip beside the loop~\cite{gargiulo2021}.
We deliver $\Phi_{\mathrm{ext}}$ with a magnetic hose~\cite{navau2014, sanchez2021},
Fig.~\ref{sfig:device}c: a superconducting aluminium cylinder wound with superconducting wire and
mounted above the tunnel. The superconducting shell guides the magnetic field of the winding to
the SNAIL loops~\cite{sanchez2021}, while a longitudinal slot cut along its side breaks the closed
path that a screening supercurrent would otherwise form around the cylinder and expel the flux.
The single-photon lifetime $T_1$ of the SNAIL oscillator across the operating flux window is
given in Table~\ref{stab:parameters} and Fig.~\ref{sfig:snailflux}c.

\begin{figure}[tbp]
  \centering
  \includegraphics[width=0.85\textwidth]{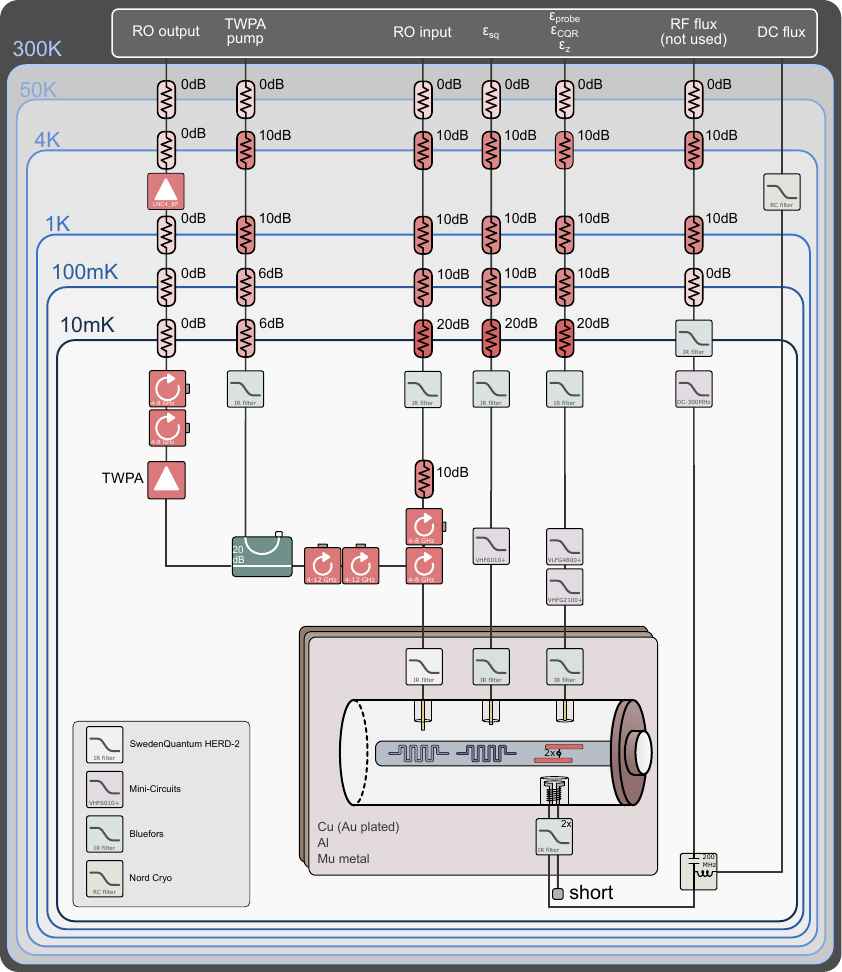}
  \caption{\textbf{Cryogenic wiring.} Attenuation, filtering, and amplification on each line of
  the dilution refrigerator, stage by stage from room temperature ($300$~K) to the mixing chamber
  ($10$~mK). Attenuators are labelled by their value at each stage. Room-temperature pulse
  generation and readout digitisation are performed with a Quantum Machines OPX1000 controller,
  and the DC flux bias is supplied by a BILT BE2142 voltage source biasing a 2 k$\Omega$ resistance placed at 4 K. The readout input, the
  squeezing-drive line ($\varepsilon_{\mathrm{sq}}$), and the combined drive line carrying the spectroscopy
  probe ($\varepsilon_{\mathrm{probe}}$), cat-quadrature readout conversion drive
  ($\varepsilon_{\mathrm{CQR}}$), and single-photon drive ($\varepsilon_x$) are attenuated and
  low-pass and infrared filtered before reaching the sample. The readout output line passes
  through two circulators, an MIT Lincoln Laboratory TWPA (pumped via a dedicated microwave line),
  and two further isolators at the mixing chamber, and is amplified again by a HEMT amplifier
  at $4$~K. The DC flux line is RC filtered at $4$~K and low-pass filtered at the mixing chamber.
  The RF flux line was installed but terminated with a short and not used in this experiment. The
  sample enclosure sits inside a nested gold-plated OFHC copper, aluminium 6061, and amumetal 4K shield stack.
  Component vendors are listed in the inset legend.}
  \label{sfig:wiring}
\end{figure}

The cryogenic wiring is shown in Fig.~\ref{sfig:wiring}. At room temperature, microwave pulses
are synthesised and digitised using a Quantum Machines OPX1000 controller, while the static coil
current setting $\Phi_{\mathrm{ext}}$ is supplied by a low-noise DC source (BILT BE2142). Three microwave
input lines reach the sample alongside the DC flux-bias line: the readout input, the squeezing-drive line
carrying $\varepsilon_{\mathrm{sq}}$, and a combined drive line carrying the spectroscopy probe
tone $\varepsilon_{\mathrm{probe}}$, the cat-quadrature readout conversion tone
$\varepsilon_{\mathrm{CQR}}$, and the single-photon drive $\varepsilon_x$. Each microwave input is
attenuated by $50$~dB distributed across the $4$~K, $1$~K, $100$~mK, and $10$~mK stages, with an
additional $10$~dB attenuator and a directional coupler on the readout input at the mixing
chamber, and passes through low-pass and infrared filters before reaching the sample. The
readout output signal passes through two $4$ to $8$~GHz circulators and a MIT Lincoln Laboratory
travelling-wave parametric amplifier (TWPA) at the mixing chamber, followed by two additional
circulators and a high-electron-mobility-transistor (HEMT) amplifier at the $4$~K stage. The
sample package is enclosed in nested gold-plated copper, aluminium, and mu-metal shields at the
mixing chamber.

\section{Circuit Hamiltonian}
\label{note:hamiltonian}

\paragraph{The circuit.} We derive the Hamiltonian of Eq.~\eqref{eq:twomode} from the
circuit. The device is a SNAIL oscillator~\cite{frattini2017},
capacitively coupled to the readout resonator, which carries the
squeezing drive. With bare modes $\hat a_0$ and $\hat b_0$, $\hat H = \hat H_a + \hat H_b + \hat H_c
+ \hat H_{\mathrm{p}}$ with
\begin{align}
  \hat H_a &= \omega_{a,0}\,\hat a_0^\dagger\hat a_0
            + g_3\big(\hat a_0 + \hat a_0^\dagger\big)^3
            + g_4\big(\hat a_0 + \hat a_0^\dagger\big)^4, \nonumber\\
  \hat H_b &= \omega_{b,0}\,\hat b_0^\dagger\hat b_0, \nonumber\\
  \hat H_c &= g\big(\hat a_0 + \hat a_0^\dagger\big)\big(\hat b_0 + \hat b_0^\dagger\big),
  \nonumber\\
  \hat H_{\mathrm{p}} &= \varepsilon_{\mathrm{sq}}\big(\hat b_0\,e^{i\omega_{\mathrm{sq}}t}
                       + \hat b_0^\dagger e^{-i\omega_{\mathrm{sq}}t}\big).
  \label{eq:circuitH}
\end{align}
Here $\omega_{a,0}$ and $\omega_{b,0}$ are the bare mode frequencies, $g_3$ and $g_4$
the cubic and quartic coefficients of the SNAIL potential, expanded about its minimum at
the operating flux in the zero-point phase of the oscillator, and $g$ a linear coupling. We
take $g_3$ and $g_4$ from the SNAIL circuit model fitted to the oscillator frequency and Kerr
coefficient measured across flux (Table~\ref{stab:parameters}).

\paragraph{Hybridisation.} The linear part of Eq.~\eqref{eq:circuitH} is diagonalised by
the normal modes of the circuit, $\hat a$ and $\hat{\tilde b}$, the modes of
Eq.~\eqref{eq:twomode}, with dressed frequencies $\omega_a$ and $\omega_b$, the ones
spectroscopy measures. Each normal mode carries a share of the phase across the SNAIL, so
the coordinate the nonlinearity acts on becomes
\begin{equation}
  \hat a_0 + \hat a_0^\dagger
  = \big(\hat a + \hat a^\dagger\big) + \lambda\big(\hat{\tilde b} + \hat{\tilde b}^\dagger\big), \qquad
  \lambda = \frac{\varphi_b}{\varphi_a} \ll 1,
  \label{eq:phasesplit}
\end{equation}
with $\varphi_a$ and $\varphi_b$ the zero-point phases of the two modes across the SNAIL.
To first order in the coupling $\lambda \approx g/(\omega_b - \omega_a)$, but at
$\omega_b \approx 2\omega_a$ the counter-rotating part of $\hat H_c$ changes this by about
a third, with a sign set by the type of coupling. We therefore never compute $\lambda$ from
$g$; it follows from the fitted $g_{21}$ (Sec.~\ref{note:twomodefit}). Through the
admixture of $\hat a$ in $\hat b_0$, the squeezing drive also drives the oscillator directly,
with amplitude $\lambda\varepsilon_{\mathrm{sq}}$ and detuning
$\omega_{\mathrm{sq}} - \omega_a \approx \omega_a$. This drive displaces the SNAIL's phase
less than the drive through the readout resonator, by the factor
$\delta_b/(\omega_{\mathrm{sq}} - \omega_a)$, a few percent, and the drive calibration
absorbs it.

\paragraph{From circuit to mode parameters.} Substituting Eq.~\eqref{eq:phasesplit} into
$\hat H_a$ and keeping the terms that survive the rotating-wave approximation in the
frames of Eq.~\eqref{eq:twomode}, the cubic term produces one coefficient and the quartic
term two,
\begin{equation}
  g_{21} = 3\,g_3\lambda, \qquad
  K_4 = -6\,g_4, \qquad
  \chi_{ab} = -24\,g_4\lambda^2 = 4K_4\lambda^2 .
  \label{eq:circuitcoeffs}
\end{equation}
The readout resonator's own self-Kerr, $K_4\lambda^4$, is smaller than the oscillator's by
$\lambda^4 \approx 10^{-7}$ and is dropped.

The Kerr coefficient $K$ of Eq.~\eqref{eq:twomode} is the one measured from the oscillator
anharmonicity. It has two contributions, the quartic term at first order and the cubic term
at second order~\cite{frattini2024prx},
\begin{equation}
  K = K_4 + \frac{30\,g_3^2}{\omega_a} = -6\,g_4 + \frac{30\,g_3^2}{\omega_a} .
  \label{eq:kerrsplit}
\end{equation}
On this device the two are comparable. The cross-Kerr of Eq.~\eqref{eq:circuitcoeffs} has
no cubic contribution and depends on $K_4$ alone, which is why we keep $K$ and $K_4$ apart.
Eq.~\eqref{eq:twomode} keeps only the resonant terms, to lowest order in $\lambda$. The
non-resonant cubic terms are dropped from it, and their effect on the oscillator is included by
using the measured $K$.

\paragraph{Purcell filter.} The filter on the readout output is a third linear mode,
coupled to the readout resonator. Including it leaves the derivation unchanged: it adds a
normal mode at $\omega_p$ with its own share $\lambda_p$ of the SNAIL's phase, and the
expansion gives an exchange $g_{21}^{(p)} = 3g_3\lambda_p$ between the SNAIL oscillator and the filter, of the same
form as $g_{21}$, and a cross-Kerr $4K_4\lambda_p^2$. We leave the filter out of
Eq.~\eqref{eq:twomode} for simplicity: its exchange is resonant only near
$\omega_{01} = \delta_p$, where we add it to the model (Sec.~\ref{note:purcell}).

\section{Calibration}
\label{note:calibration}

In this section we describe how the two-photon drive $\varepsilon_2$ and the photon number
$\bar n$ are obtained at each operating point, and how the device parameters they rest on are
fixed.

\subsection{Two-mode fit of the intra-well spectrum}
\label{note:twomodefit}

Most numbers in this work come from one fit. We fit the excited levels measured in each
spectroscopy map, such as that of Fig.~\ref{fig:spectroscopy}a of the main text, with the
two-mode model of Eq.~\eqref{eq:twomode} (Methods), derived in Sec.~\ref{note:hamiltonian}. The
fit has two free parameters: the exchange $g_{21}$, and the drive calibration
$c = \varepsilon_2/A$, the two-photon drive per volt of squeeze amplitude $A$. At each amplitude
it returns $\varepsilon_2$, $\bar n$, and the intra-well level spacings $\omega_{n-1,n}$, which
locate the resonances marked in Fig.~\ref{fig:lifetime}a of the main text.

\paragraph{Model.} The Hamiltonian solved is Eq.~\eqref{eq:twomode} (Methods) with the readout
resonator displaced by its mean field $\beta = \varepsilon_2/g_{21}$, taken real by choosing the
phases of $\hat a$ and $\hat b$,
\begin{equation}
  \hat H = \hat H_{\mathrm{KCQ}}
  - \delta_b\,\hat b^\dagger\hat b
  + g_{21}\big(\hat a^2\hat b^\dagger + \hat a^{\dagger 2}\hat b\big)
  - \chi_{ab}\,\hat a^\dagger\hat a\,\hat b^\dagger\hat b
  - \chi_{ab}\beta\,\hat a^\dagger\hat a\big(\hat b + \hat b^\dagger\big),
  \label{eq:twomodedisplaced}
\end{equation}
with $\hat H_{\mathrm{KCQ}}$ the Kerr-cat Hamiltonian of Eq.~\eqref{eq:kcq} (Methods). It keeps
the resonant terms of Sec.~\ref{note:hamiltonian} and drops the readout resonator's self-Kerr and
all higher orders. It also drops the readout linewidth, which is far below the exchange
splittings the map resolves, so it broadens the lines without moving them measurably. The
displacement produces the Stark shift of the oscillator inside $\hat H_{\mathrm{KCQ}}$ and the
residual vertex, the last term. Both carry $\chi_{ab}$ multiplied by the readout field, large
under drive, so $\chi_{ab}$ is kept although it is far below $K$.

The probe couples to $\hat a + \hat a^\dagger$, which flips the oscillator's photon-number parity, a
symmetry of Eq.~\eqref{eq:twomodedisplaced}. The lines it can excite are therefore transitions
from the ground state to levels of the opposite parity, each with a strength set by its matrix
element.

\paragraph{Fixed inputs.} The mode frequencies $\omega_a$ and $\omega_b$, and with them $\Delta$,
$\delta_b$, and the mode detuning $L$, are taken from spectroscopy at the flux point of the
measurement, and the Kerr coefficient $K$ from the oscillator anharmonicity measured there. The
cross-Kerr is not left free. Exchange and cross-Kerr both follow from the readout resonator's
participation $\lambda$ in the SNAIL's nonlinearity, as $g_{21} = 3g_3\lambda$ and
$\chi_{ab} = 4K_4\lambda^2$ (Eq.~\eqref{eq:circuitcoeffs}), so the fit takes $g_{21}$ from the
map, where the splittings of the avoided crossings fix it nearly independently of the drive
calibration, and sets
\begin{equation}
  \chi_{ab} = \frac{4K_4\,g_{21}^2}{9g_3^2}, \qquad K_4 = K - \frac{30\,g_3^2}{\omega_a} .
  \label{eq:lambdatie}
\end{equation}
The cubic coefficient enters only here. We use a single value of $g_3$ at every flux bias,
within the uncertainty of the SNAIL circuit fit of Fig.~\ref{sfig:snailflux}d across the window.
Using the fitted $g_3$ of each flux bias instead changes $\bar n$ by less than $0.3\%$
(Sec.~\ref{note:nbarerror}).

\paragraph{The fit.} The squeezing drive reaches the SNAIL oscillator through the readout resonator,
which responds linearly, so $\varepsilon_2$ is proportional to the squeeze amplitude,
$\varepsilon_2 = cA$. In each spectroscopy map, one value of $c$ fits the lines at all squeeze
amplitudes. Figure~\ref{sfig:drivecal} checks how $c$ changes with the squeezing frequency
against a simulation of the transmission through the readout resonator.

Each row of a map is a probe sweep at one squeeze amplitude, and every level the probe excites
appears as a dip in $\langle Z\rangle$. We find the dips in each row and keep
those deeper than both a fixed threshold and six times the noise of the row. We discard rows
whose $\langle Z\rangle$ is already below $0.3$ without the probe, and a $\pm 2$\,MHz band around the
drive artefact at $\omega_{\mathrm{sq}}/2$. For each trial pair $(c, g_{21})$ we compute the
model's lines at every amplitude, and the cost of the fit is the mean distance from each dip to
the nearest visible line. A line is visible if its strength, weighted by the probe power at its
frequency, is at least $3\%$ of that of the strongest line. Each distance is capped at
$8$\,MHz, so a dip the model cannot explain, such as a spur or a third mode, adds a fixed
penalty instead of dragging the fit towards it. We search $g_{21}$ between $0.3$ and $1.2$\,MHz.

We fix the other parameters because the map cannot separate them from the free ones. The level
ladder depends mainly on $\varepsilon_2/K$ and $\Delta/2K$, so $c$ and $\Delta$ can compensate
each other, and a free $\Delta$ drifts to values the lifetime map rules out. Likewise $g_{21}$
and $\chi_{ab}$ can compensate each other, since both the exchange and the Stark shift bend the
ladder as the drive grows; Eq.~\eqref{eq:lambdatie} removes this freedom. An error in $K$ is
absorbed into $c$ at the level of a few percent.

\paragraph{What the fit returns.} The photon number is the oscillator occupation of the ground state
of Eq.~\eqref{eq:twomodedisplaced}, $\bar n = \langle\hat a^\dagger\hat a\rangle$, at each
amplitude. It is not the mean field $\varepsilon_2/K + \Delta/2K$ of Eq.~\eqref{eq:nbar}. The
Stark shift lowers the effective detuning as $\varepsilon_2^2$ and the hybridisation with the
readout resonator redistributes weight, both growing with drive, so the mean-field estimate overestimates $\bar n$ at the strongest drives by an amount not small against the spread of $\bar n$ across the
lifetime map. The same solution gives the frequencies that locate the conditions
$\omega_{n-1,n} = \delta_b$ marked in Fig.~\ref{fig:lifetime}a of the main text, so the crossing
and the photon number quoted at it come from one diagonalisation. The residual is dominated by
the branch hybridising with the two-readout-photon state, whose lines lie slightly high at the
largest amplitudes; this is where a term beyond Eq.~\eqref{eq:twomodedisplaced} would first
appear, the readout resonator's own nonlinearity or the sextic term of the element, and it does
not touch the lower rungs from which $\bar n$ is read.

The photon number at the crossing hardly depends on $g_{21}$. The crossing sits where the first
gap of the Kerr oscillator, including its Stark shift, equals $\delta_b$, and that Stark shift,
$4K_4\varepsilon_2^2/9g_3^2$ by Eq.~\eqref{eq:lambdatie}, does not contain $g_{21}$. The exchange
changes $\bar n$ only through the small admixture of the readout resonator in the ground state:
moving $g_{21}$ from $0.6$ to $1.2$\,MHz changes $\bar n$ at the crossing by less than $3\%$ at
the operating points of Fig.~\ref{fig:flux} of the main text.

\subsection{Uncertainty of the photon number}
\label{note:nbarerror}

The uncertainty of $\bar n$ is set almost entirely by the Kerr coefficient, which the fit takes
as measured and cannot determine from the map: changing $K$ by $\pm 5\%$ leaves the residual of
the fit unchanged, while $\bar n$, which at the readout resonance scales as $L/K$
(Eq.~\eqref{eq:nmin}), moves by nearly the same fraction.

Each operating point is assigned its own relative uncertainty of $K$: the larger of $5\%$, about
1.4 times the rms deviation of single calibrations of $K$ from the circuit fit, and the relative
difference between the $K$ used for $\bar n$ and a reference value. The reference is the mean of
the calibrations at that flux bias in Fig.~\ref{sfig:snailflux}b, or the circuit fit where there
are none. We refit the map of each operating point with $K$ shifted by this uncertainty in both
directions, with $c$ and $g_{21}$ free, and evaluate $\bar n$ at the drive at which the lifetime
was measured. These are the error bars on $\bar n$ in Fig.~\ref{fig:flux} of the main text.

The other inputs of the fit contribute less. An error of $\pm 0.2$\,MHz in the oscillator
frequency, and with it in $\Delta$, moves $\bar n$ by less than $0.6\%$. Moving $g_{21}$ from
$0.6$ to $1.2$\,MHz changes it by less than $3\%$ (Sec.~\ref{note:twomodefit}), and using the
fitted $g_3$ of each flux bias by less than $0.3\%$. The model omits the part of the Stark shift
that the cubic term produces off resonance; including it lowers $\bar n$ by less than $2\%$.

\subsection{Drive calibration against the simulated transmission}
\label{note:crosschecks}

\begin{figure}[tbp]
  \centering
  \includegraphics{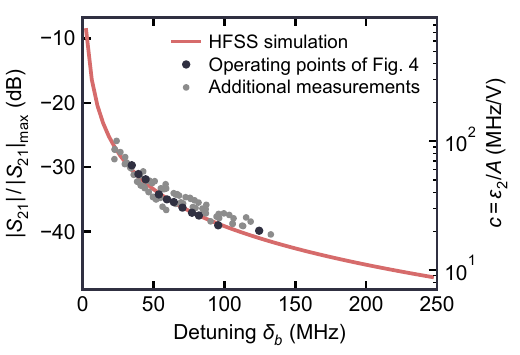}
  \caption{\textbf{Drive calibration against the simulated transmission of the readout
  resonator.} Two-photon drive per volt of squeeze amplitude, $c = \varepsilon_2/A$ (dots,
  right axis), from the two-mode fit of excited-level spectroscopy maps of the kind shown in
  Fig.~\ref{fig:spectroscopy}a of the main text (Sec.~\ref{note:twomodefit}), against the
  squeezing-readout detuning $\delta_b$ of each map. Black dots are the operating points of
  Fig.~\ref{fig:flux} of the main text, and grey dots further maps of the flux campaign. Fits
  that failed, with $c$ unconstrained or the dips off the model lines, are not shown. The red
  line is an HFSS simulation of $|S_{21}|$, the transmission from the pin under the readout
  resonator to the SNAIL, normalised to its maximum (left axis), with $\delta_b$ measured from
  the simulated readout resonance. One decade of the right axis spans $20$\,dB of the left, so
  the two axes differ by one overall factor, the only free parameter of the comparison.}
  \label{sfig:drivecal}
\end{figure}

The squeezing drive reaches
the SNAIL through the readout resonator. The drive calibration $c$ should therefore be largest
near the readout resonance and fall as $1/\delta_b$ away from it, modified by the rest of the
drive line, which an HFSS simulation of the transmission $|S_{21}|$ from the pin under the
readout resonator (Fig.~\ref{sfig:device}a) to the SNAIL includes. Since the fit obtains $c$ from
the spectroscopy alone, the simulation is an independent check. Figure~\ref{sfig:drivecal}
compares the two, at $\delta_b/2\pi$ from $22$ to $133$\,MHz. The simulation gives the frequency
dependence of the transmission but not the absolute scale of $c$, so the comparison leaves one
overall factor free. With that factor set, the fitted $c$ follows the simulated curve over a
range in which it changes by a factor of five.

\section{Bit-flip time from click records}
\label{note:clickrecord}

In this section we use click records, continuous sequences of repeated cat-quadrature
readouts that monitor the qubit, to measure the bit-flip time and the spectrum of the
readout signal. The extracted values correspond to the bit-flip times of
Fig.~\ref{fig:control}d from the main text; the fitting procedure that extracts the flip probability per
readout is described in the Methods.

For this experiment we work at the autoparametric resonance with the Purcell filter ($\bar n = 33.8$). The cat
never leaves the operating point. Each record follows the following  sequence: the two-photon drive
ramps from zero to the operating amplitude in $4$ to $16~\mu$s, which prepares the cat in a
random well, and then stays on at fixed amplitude for the whole record, up to $1000$~s. From $100~\mu$s after the ramp, a $4~\mu$s
cat-quadrature readout repeats at the same drive amplitude, with a readout interval $D$. The readout
drive is switched off between readouts, except for back-to-back readout at $D = 4~\mu$s. No
ramp is applied between the hold and the readouts, so every flip happens at the operating point,
under the two-photon drive and the repeated readout.

\subsection{Counting bit flips}

Each readout yields the quadrature \(I\) along the axis connecting the two readout lobes (see Methods). From a calibration performed at the same settings immediately before the run, we centre this axis at the midpoint of the two lobe centres and define \(h\) as half their separation; the calibrated lobe centres are therefore located at $I = \pm h$. The calibration shots are each prepared anew,
so both lobes are populated even when a record stays in one well, and the calibration is
repeated when the single-shot signal-to-noise ratio or the lobe positions drift. A readout
is assigned to $|{+Z}\rangle$ once $I > h/2$ and to $|{-Z}\rangle$ once $I < -h/2$, and
readouts in between keep the previous assignment. A flip is a change of the assigned
state, counted on single readouts without averaging. The first $1$~ms of every record,
when the readout drive switches on, is excluded from both the flip count and the exposure.

\begin{figure}[tbp]
  \centering
  \includegraphics[width=\textwidth]{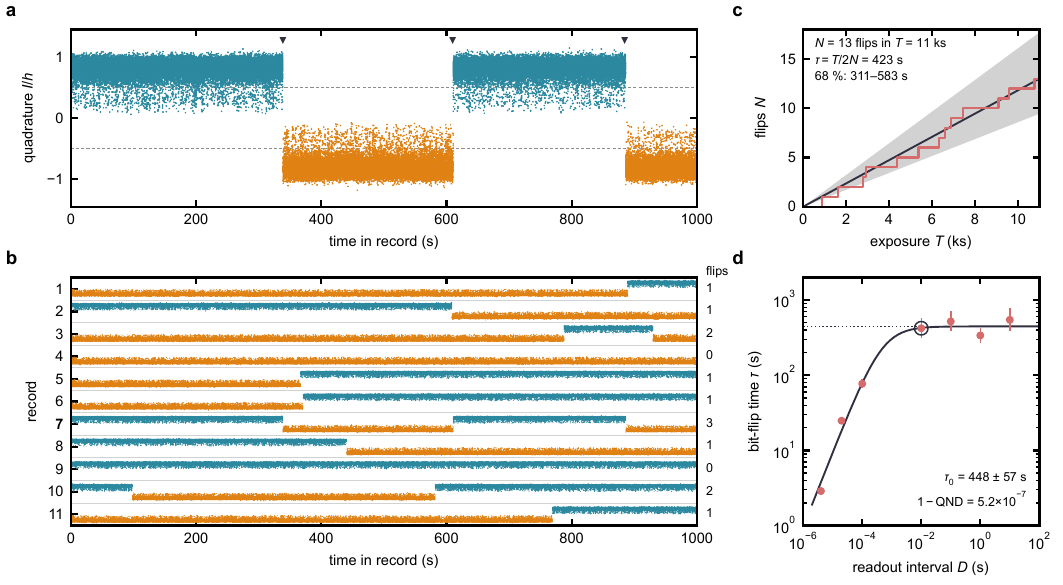}
  \caption{\textbf{Bit-flip time from click records.}
  \textbf{a,} One record of a run with a readout every $10$~ms: quadrature $I/h$ of every
  readout, coloured by the assigned state. Dashed lines mark the assignment
  thresholds at $\pm h/2$, and triangles mark the flips.
  \textbf{b,} Every record of the run, plotted as in \textbf{a}, with the number of flips
  in each record at right.
  \textbf{c,} Cumulative number of flips against exposure, with the records placed end to
  end. The solid line is $N = T/2T_X$, and the shaded band is its 68\% Poisson interval.
  \textbf{d,} Bit-flip time against the readout interval $D$, from all runs at this point;
  the circled point is the run of \textbf{a}–\textbf{c}. The solid line is the
  fit of $1/T_X = 1/T_X^0 + 2p_{\mathrm{flip}}/D$ described in Methods, and the dotted line
  marks $T_X^0$. Error bars are 68\% Poisson intervals on the flip counts.}
  \label{sfig:clickrecord}
\end{figure}

For a symmetric telegraph signal whose flips occur at a rate $1/(2T_X)$, the correlation
between outcomes at time $t$ apart decays as $e^{-t/T_X}$, the same decay that defines every
other bit-flip time in this work. $N$ flips in a total exposure $T$ therefore give
$T_X = T/2N$, and we quote exact 68\% Poisson intervals on $N$.
Fig.~\ref{sfig:clickrecord}a–c shows one run with $D = 10$~ms: eleven records of
$1000$~s contain $13$ flips, which gives $T_X = 423$~s ($68\%$ interval $311$ to
$583$~s). The mean times the cat spends in $|{+Z}\rangle$ and $|{-Z}\rangle$ before flipping,
$821$ and $867$~s ($6$ and $7$ flips), agree within their uncertainty, consistent with equal
flip rates out of the two wells.

\subsection{Spectrum of the readout signal}
\label{note:clickspectrum}

\begin{figure}[tbp]
  \centering
  \includegraphics[width=\textwidth]{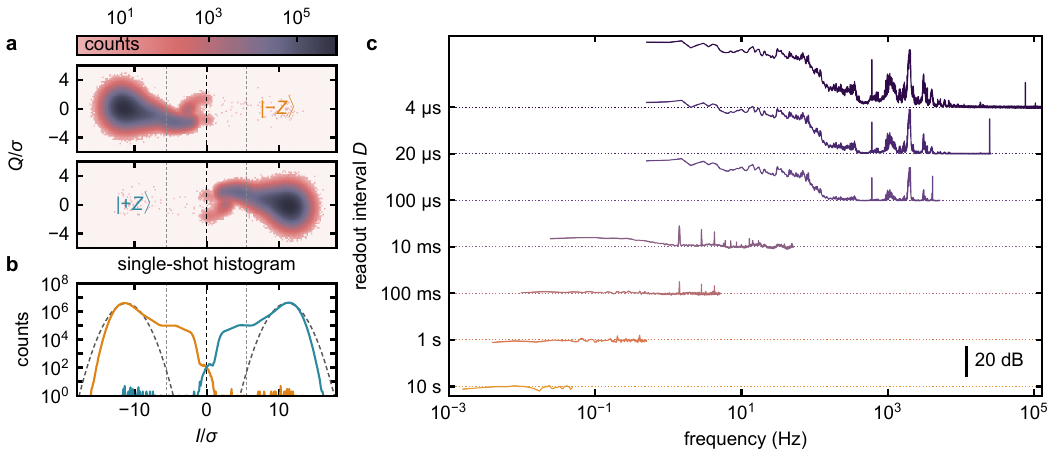}
  \caption{\textbf{Single shots and spectrum of the readout signal in the click records.}
  \textbf{a,} Every readout of the back-to-back record ($D = 4~\mu$s) of Fig.~\ref{fig:control}c of the main
  text, sorted by the sign of $I$ in the preceding readout: $|{-Z}\rangle$ after $I < 0$
  (top) and $|{+Z}\rangle$ after $I > 0$ (bottom). Quadratures are in units of the noise
  width $\sigma$, and counts are on a logarithmic scale. The black dashed line marks $I = 0$,
  and the grey dashed lines mark the thresholds at $I = \pm h/2$.
  \textbf{b,} The two sets of \textbf{a} along $I/\sigma$, $|{-Z}\rangle$ in orange and
  $|{+Z}\rangle$ in blue. Grey dashed curves are Gaussians centred on the two lobes with
  the measured lobe width, and the vertical lines are as in \textbf{a}.
  \textbf{c,} Power spectral density of the quadrature signal for each readout interval
  $D$, in decibels above its white noise floor (dotted lines). The spectra are offset
  vertically, and each extends to its Nyquist frequency $1/2D$.}
  \label{sfig:readoutspectrum}
\end{figure}

Fig.~\ref{sfig:readoutspectrum}a,b shows the single shots of the back-to-back record of
Fig.~\ref{fig:control}c of the main text. Each readout is sorted by
the sign of $I$ in the preceding readout, $I < 0$ heralding $|{-Z}\rangle$ and $I > 0$
heralding $|{+Z}\rangle$. This sorting keeps every readout, whereas the assignment
infidelity (Methods) uses only heralds with $|I| > h/2$. Of the readouts that land on the
wrong side of $I = 0$, nearly nine in ten lie in the tail between $I = 0$ and the
threshold at $h/2$; the rest sit at the opposite lobe and are bit flips.

Fig.~\ref{sfig:readoutspectrum}c shows the power spectral density of the quadrature
signal in click records with $D$ from $4~\mu$s to $10$~s. The quadrature signal is
$I$ multiplied by the sign of the assigned state, so that both lobes sit at $+h$ and the
flips drop out of the spectrum. Spectra taken at different $D$ agree once aliasing is
accounted for, so the features do not appear to originate in the readout itself. We
suspect three sources: vibrations of the cryostat excited by the pulse-tube cryocooler,
which shape the spectrum from low frequencies to a few
kilohertz; a weakly damped mechanical resonance of the chip, which gives a narrow line
with harmonics; and spurious pickup in the readout chain, at the highest frequencies. The
strongest feature is a series of bursts in which the quadrature signal drops towards zero, and
these bursts likely produce the tail between the two lobes in
Fig.~\ref{sfig:readoutspectrum}a,b. Since the readout noise shrinks together with the signal during a burst, the bursts could
come from fluctuations of the amplifier gain. We see no correlation between the bit flips and the
bursts, and we do not expect these features to limit the results reported here.

\section{Measurement of the phase-flip time}
\label{note:phaseflipgates}

In this section we describe the Ramsey sequence behind the phase-flip time $T_Z$ and its two
$X(\pi/2)$ gates: the Kerr refocusing gate~\cite{grimm2020, kirchmair2013} and a phase-modulated gate, a single-lobe version of a
previously demonstrated detuning gate~\cite{hajr2024}, with which we also observe Rabi oscillations under a
$Z$ drive.

The sequence (Fig.~\ref{sfig:phaseflipgates}b,e) heralds a well state with a first cat-quadrature
readout and applies an \(X(\pi/2)\) gate at the gate point \(G\), preparing an equal superposition of the two logical states. If the gate point \(G\) differs from the operating point \(H\), the two-photon-drive amplitude is then ramped from \(G\) to \(H\) in 48–64 ns. The qubit evolves at \(H\) for a variable hold time \(t\), after which the drive is ramped back to \(G\). A second \(X(\pi/2)\) gate converts the remaining coherence into a difference between the two well populations, which is measured by a second readout. Phase flips during the hold destroy the coherence of the superposition, causing \(\langle Y\rangle\) to decay towards zero. Fitting the decay of \(\langle Y\rangle(t)\) (Fig.~\ref{sfig:phaseflipgates}c,f) yields \(T_Z\).

\subsection{Kerr refocusing gate}

For the Kerr refocusing gate, the two-photon drive is switched off and the oscillator evolves under its
Kerr nonlinearity alone~\cite{grimm2020, kirchmair2013}. Because $n(n-1)$ is even for every Fock state $|n\rangle$, the unitary
$e^{iK\hat a^{\dagger 2}\hat a^2 t}$ returns every Fock component to its initial phase at
$t = \pi/K$, refocusing a coherent state onto itself. Halfway through this evolution, at $t = \pi/2K$, the
state is an equal superposition of two coherent states of opposite phase, which the two-photon drive
captures as an $X(\pi/2)$ rotation when switched back on with its phase matched to the free
precession of the mode during the gate. Scanning the drive-off duration displays the
revival (Fig.~\ref{sfig:phaseflipgates}a); its period fixes the gate time and the
recapture phase at each gate point.

Although the Kerr refocusing gate requires no shaped waveform, the cat retains its full photon number throughout the duration
$\pi/2K$, so the gate error from single-photon loss grows linearly with $\bar n$. Higher-order
Kerr terms also degrade the refocusing, as the phases they accumulate are no longer integer multiples
of $2\pi$ at $t = \pi/K$.

We therefore execute the gate at a lower photon number than the target operating point. The squeezing drive
ramps from the readout plateau $M$ down to the gate point $G$, where the gate is applied,
and then ramps to $H$ for the hold (Fig.~\ref{sfig:phaseflipgates}b). Because the gate duration
$\pi/2K$ is set by the Kerr nonlinearity rather than the drive amplitude, decoupling the gate point from the hold point allows us to
measure phase-flip times at $H$ that are shorter than the gate itself.

At large photon numbers the Ramsey contrast falls, because much of the coherence is lost
before the hold begins. A large part is lost during the gates, while the squeezing drive is
off, through the high single-photon loss of the oscillator. Further coherence is lost during
the ramps between $G$ and $H$. Their duration is fixed, while $T_Z$ at $H$ shortens as
$1/\bar n$, so this loss grows with every added photon. As the contrast falls, each step to
higher $\bar n$ requires much more averaging to resolve the decay, which sets the largest
photon number in Fig.~\ref{fig:flux}b of the main text.

\subsection{Phase-modulated gate}
\label{note:chirpgate}

An $X(\pi/2)$ rotation can also be realised without switching off the squeezing drive by modulating only the phase
$\phi_{\mathrm{sq}}(t)$ of the two-photon drive~\cite{hajr2024}. In the frame rotating at $\phi_{\mathrm{sq}}(t)/2$, the phase modulation acts as an effective
detuning pulse, $\Delta_{\mathrm{eff}}(t) = \Delta - \dot\phi_{\mathrm{sq}}(t)/2$. The rotation rate about
the logical $X$ axis is given by the tunnel splitting of the ground doublet of the Kerr-cat
Hamiltonian at $\Delta_{\mathrm{eff}}(t)$; integrated over the pulse, this splitting generates the gate rotation. Crucially, the splitting opens appreciably only where $\Delta_{\mathrm{eff}} < 0$, where the double-well barrier drops and the wells move closer together. The original detuning gate uses a zero-area phase profile with a second excursion of opposite sign to return
$\phi_{\mathrm{sq}}$ to zero~\cite{hajr2024}. However, that second lobe contributes no rotation—it instead swells the cat and accumulates single-photon loss for as long as the active half of the gate.

We retain only the barrier-lowering lobe. While the two-photon drive amplitude is held constant, its phase
follows
\begin{align}
  \phi_{\mathrm{sq}}(t) &= \Delta_{\mathrm{ch}}\left[t - \frac{t_{\mathrm{g}}}{2\pi}
             \sin\!\left(\frac{2\pi t}{t_{\mathrm{g}}}\right)\right],
  \qquad 0 \le t \le t_{\mathrm{g}},
  \label{eq:chirpphase}
\end{align}
producing a single detuning lobe $\Delta_{\mathrm{eff}}(t) = \Delta - \Delta_{\mathrm{ch}}\sin^2(\pi t/t_{\mathrm{g}})$
of depth $\Delta_{\mathrm{ch}}$ (peak drive-frequency excursion $\dot\phi_{\mathrm{sq},\max} = 2\Delta_{\mathrm{ch}}$) and duration $t_{\mathrm{g}}$. Over the course of the gate, the drive accumulates a net phase advance of \(\phi_{\mathrm{end}}=\Delta_{\mathrm{ch}}t_{\mathrm{g}}\).
Rather than undoing this phase step with a second lobe, we account for it through a virtual frame update: every subsequent
two-photon drive pulse is advanced in phase by $\phi_{\mathrm{end}}$, and the
cat-quadrature readout—which references the oscillator field at half the squeezing frequency—is advanced by
$\phi_{\mathrm{end}}/2$. Because $\phi_{\mathrm{end}}$ is fixed analytically by the programmed waveform, it is computed directly
rather than fitted.

By simulating with a single-mode master-equation using the measured oscillator $T_1$ and a thermal occupation $n_{\mathrm{th}} = 0.1$, this single-lobe gate incurs two to three times
less error than the original detuning gate at the same photon number, and its advantage over the
Kerr refocusing gate grows with $\bar n$ because the wells contain only a few photons while the state
rotates. This comes with a trade-off on the drive bandwidth: the peak excursion of the squeezing frequency,
$2\Delta_{\mathrm{ch}}$, scales nearly linearly with $\bar n$, requiring the squeezing-drive line to maintain a flat amplitude and phase response across that span.

We calibrate the phase-modulated gate by inserting a pair of gates between two readouts, so that two consecutive
$X(\pi/2)$ rotations compose an $X(\pi)$ gate that switches the well. An initial scan over
$\Delta_{\mathrm{ch}}$ and the readout phase fixes both the sign of the frequency
excursion—as the opposite sign swells the cat without inducing switching in the
readout reference frame. A chevron scan over $\Delta_{\mathrm{ch}}$ and $t_{\mathrm{g}}$ then selects the shortest gate
on the first switching fringe, after which a train of $4k+2$ gates refines $\Delta_{\mathrm{ch}}$ to an
exact $\pi/2$ rotation. Because the squeezing drive remains on throughout, the gate is calibrated and operated directly at
the target point ($G = H$), eliminating the amplitude ramps between the gates and the hold
(Fig.~\ref{sfig:phaseflipgates}e).

Although the phase-modulated gate gives a higher signal-to-noise ratio, its Ramsey signal shows a
beating, which we attribute to leakage into the excited doublets during the gate. We
therefore chose the Kerr refocusing gate for the phase-flip times reported in this work,
both for the $T_Z$ scaling across flux biases in Fig.~\ref{fig:flux}b of the main text and
for the readout-resonance point ($T_Z = 208 \pm 40$~ns at $\bar n = 22.4$). At the $\bar n = 13.5$
autoparametric resonance of Fig.~\ref{fig:flux}, we measure $T_Z$ with both gates
(Fig.~\ref{sfig:phaseflipgates}c,f): the Kerr refocusing gate played at a lower squeezing
amplitude ($G \neq H$) yields $T_Z = 354 \pm 45$~ns, and the phase-modulated gate operated
directly at the resonance ($G = H$) yields $T_Z = 351 \pm 67$~ns.

Although the phase-modulated gate avoids amplitude ramps, its Ramsey contrast also decreases at large photon numbers. The detuning lowers the barrier and crosses resonances that can excite the oscillator out of its ground doublet, such as those shown in Fig. S11. We attribute the beating observed in the Ramsey signal to leakage induced by these crossings. Moreover, during the barrier-lowering part of the pulse, the cat temporarily occupies a regime with weaker bit-flip protection. Both effects become more pronounced as the photon number increases. 

\subsection{$Z$-gate Rabi oscillations}
\label{note:zrabi}

\begin{figure}[tbp]
  \centering
  \includegraphics[width=\textwidth]{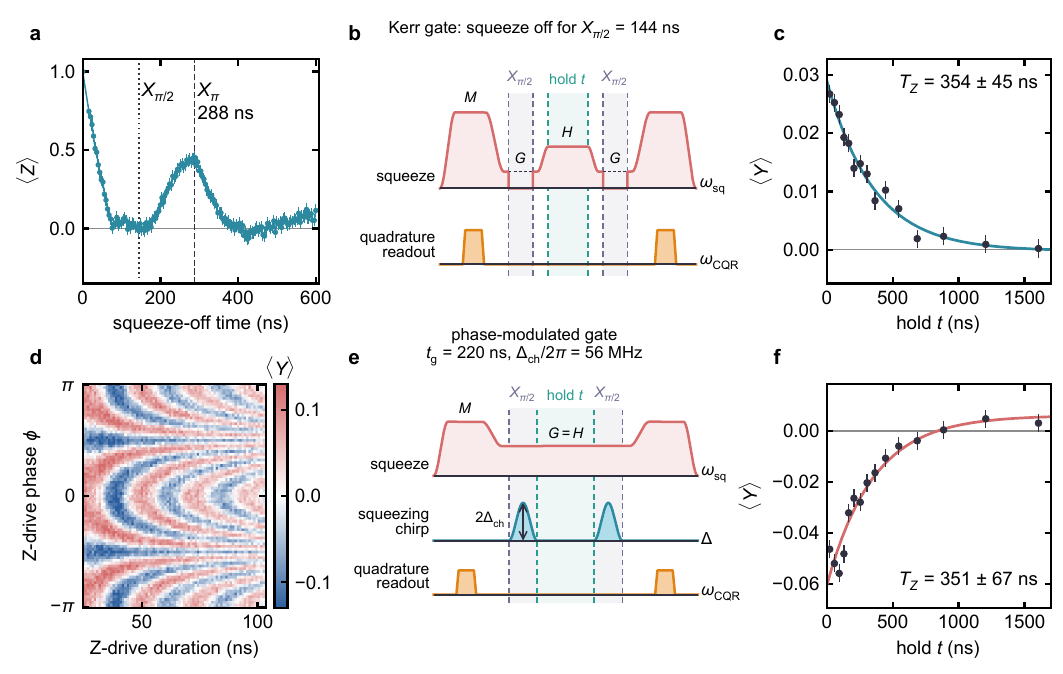}
  \caption{\textbf{Coherent control and phase-flip time with the two $X(\pi/2)$ gates.}
  Kerr refocusing gate (top row) and phase-modulated gate (bottom row), both at the
  autoparametric resonance at $\bar n = 13.5$.
  \textbf{a,} Kerr revival at the gate point $G$: expectation value $\langle Z\rangle$ over its value without the gate, against the drive-off duration. Dotted and
  dashed lines mark the $X(\pi/2)$ and $X(\pi)$ gate times.
  \textbf{b,} Ramsey sequence with the Kerr refocusing gate, showing the two-photon drive
  (top) and the readout (bottom). The gates are played at $G$ with the drive switched off,
  and the hold of duration $t$ takes place at $H$; $M$ is the readout plateau.
  \textbf{c,} Ramsey signal $\langle Y\rangle$ against hold time $t$ with the Kerr refocusing
  gate. The solid line is a fit to $A e^{-t/T_Z} + B$.
  \textbf{d,} $Z$-gate Rabi oscillations with the phase-modulated gates: $\langle Y\rangle$ against the
  phase $\phi$ and duration of a single-photon $Z$ drive at $\omega_{\mathrm{sq}}/2$ applied
  between the two gates, with $\phi$ measured from the fitted $\phi_0$.
  \textbf{e,} Ramsey sequence with the phase-modulated gate. The drive amplitude stays constant
  ($G = H$), and each gate is a single detuning lobe of depth $\Delta_{\mathrm{ch}}$ and
  duration $t_{\mathrm{g}}$ (Eq.~\eqref{eq:chirpphase}).
  \textbf{f,} As in \textbf{c}, with the phase-modulated gate.
  Error bars are one standard deviation, estimated from the point-to-point scatter of the
  data.}
  \label{sfig:phaseflipgates}
\end{figure}

Once calibrated, the pair of phase-modulated \(X(\pi/2)\) gates maps logical-\(Z\) rotations onto interwell switching. A resonant single-photon \(Z\) drive at \(\omega_{\mathrm{sq}}/2\), applied between the gates with phase \(\phi\) for a duration \(\Delta t\), produces a rotation
\[
\theta=\Omega\Delta t\cos(\phi-\phi_0).
\]
The resulting \(Z\)-gate Rabi fringes are shown in Fig.~\ref{sfig:phaseflipgates}d at \(\bar n=13.5\) and in Fig.~\ref{fig:control}a at \(\bar n=22.4\), where \(t_{\mathrm g}=204~\mathrm{ns}\) and \(\Delta_{\mathrm{ch}}/2\pi=135.25~\mathrm{MHz}\).

\begin{figure}[tbp]
  \centering
  \includegraphics[width=\textwidth]{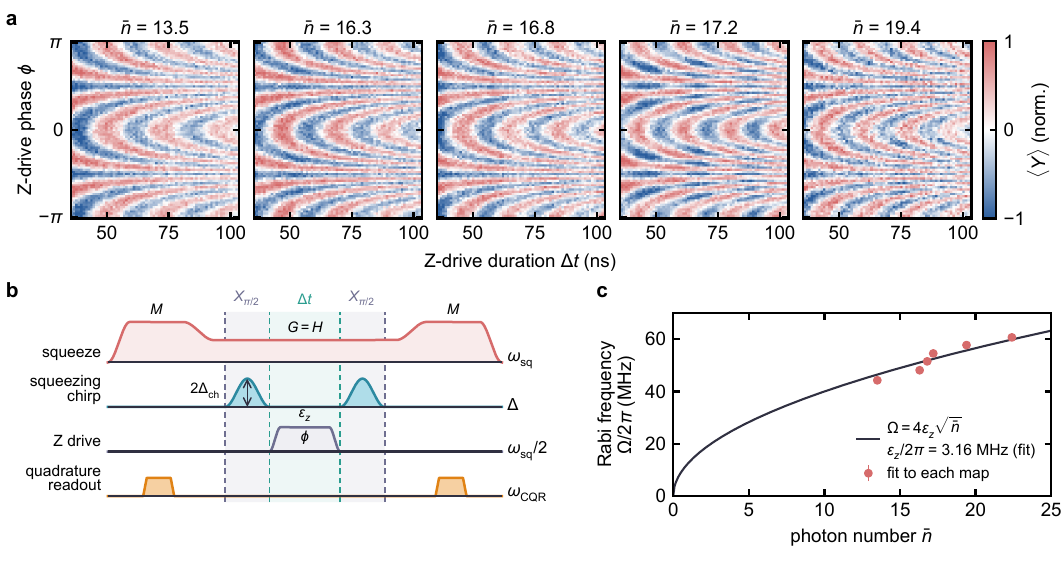}
  \caption{\textbf{$Z$-gate Rabi oscillations against photon number.}
  \textbf{a,} Expectation value $\langle Y\rangle$ against the duration
  $\Delta t$ and phase $\phi$ of the $Z$ drive, at several of the photon numbers of
  Fig.~\ref{fig:flux} of the main text and the same drive amplitude, with $\phi$ measured from
  each map's fitted $\phi_0$. Each map is normalised to its own contrast.
  \textbf{b,} Pulse sequence, not to scale. The phase-modulated $X(\pi/2)$ gates and the $Z$ drive
  at $\omega_{\mathrm{sq}}/2$ are played at constant squeezing amplitude ($G = H$); $M$ is the
  readout plateau.
  \textbf{c,} Rabi frequency $\Omega$ from a global fit to each map against $\bar n$,
  including $\bar n = 22.4$, whose map is not shown. The solid line is a fit to
  $\Omega = 4\varepsilon_z\sqrt{\bar n}$ with $\varepsilon_z$ the only free parameter.
  Error bars are one standard error of the fit, smaller than the markers.}
  \label{sfig:catphase2d}
\end{figure}

We repeat this measurement for several photon-number values of Fig.~\ref{fig:flux} of the main text
with the same $Z$-drive amplitude $\varepsilon_z$ (Fig.~\ref{sfig:catphase2d}). The Rabi
frequency grows as $\Omega = 4\varepsilon_z\sqrt{\bar n}$, as expected for a single-photon
drive, which shifts the energies of the two wells in opposite directions in proportion to
the cat amplitude $\alpha = \sqrt{\bar n}$.

\section{Effective temperature of the driven oscillator}
\label{note:thermometry}

In this section we measure the steady-state population of the first excited manifold at an amplitude
drive where the autoparametric cooling is inactive. The ratio to the ground-manifold population estimates
the heating ratio $\Gamma_\uparrow/\Gamma_\downarrow$ of Sec.~\ref{note:floor}. To do so 
we follow the incoherent spectroscopy experiment of Ref.~\cite{adinolfi2025} at the flux bias and squeezing frequency of Fig.~\ref{fig:spectroscopy} of the main text, for a
squeeze amplitude $A = 0.999$~V, above the autoparametric band.

\begin{figure}[tbp]
  \centering
  \includegraphics[width=\textwidth]{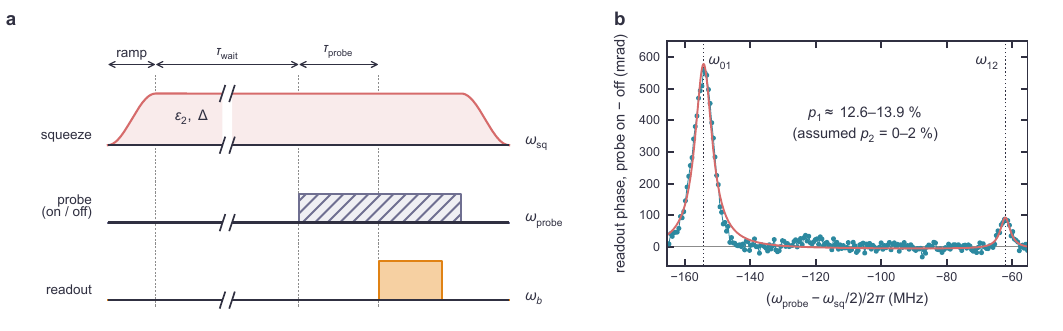}
  \caption{\textbf{Incoherent spectroscopy of the driven oscillator.} \textbf{a,} Pulse
  sequence, not to scale. The squeezing drive at \(\omega_{\mathrm{sq}}\) is ramped in amplitude and detuning to set \(\varepsilon_2\) and \(\Delta\), and is then held for \(\tau_{\mathrm{wait}}=30~\mu\mathrm{s}\). A weak probe at \(\omega_{\mathrm{probe}}\) is subsequently applied for \(\tau_{\mathrm{probe}}=10~\mu\mathrm{s}\) and remains on during the readout at \(\omega_b\). Measurements with and without the probe are interleaved (hatched region).
\(\mathbf{b,}\) Probe-induced change in the readout phase, plotted as a function of the probe detuning from \(\omega_{\mathrm{sq}}/2\). Each point averages 4000 repetitions. The solid line is
  a fit of two Lorentzians on a common offset, and dotted lines mark the transitions at
  $\omega_{01}$ and $\omega_{12}$. The population $p_1$ of the first excited manifold
  written in the panel assumes a population $p_2 = 0$ to $2\%$ of the second. Measured at
  $A = 0.999$~V, at the flux bias and squeezing frequency of Fig.~\ref{fig:spectroscopy} of the
  main text.}
  \label{sfig:thermometry}
\end{figure}

The pulse sequence is shown in Fig.~\ref{sfig:thermometry}a. The squeezing drive is switched on with
a $4~\mu$s raised-cosine ramp of its amplitude, during which its frequency is swept from
$2\omega_a$ to $\omega_{\mathrm{sq}}$, so that the detuning rises adiabatically from zero
to $\Delta/2\pi = 65.3$~MHz, with $\omega_a/2\pi = 3.7606$~GHz. A wait of
$\tau_{\mathrm{wait}} = 30~\mu$s, about nine times the relaxation time of the first
excited manifold at this drive, $T_1 = 3.26 \pm 0.06~\mu$s, brings the oscillator to its
steady state. A weak probe at $\omega_{\mathrm{probe}}$ is then applied for
$\tau_{\mathrm{probe}} = 10~\mu$s and held through a $4~\mu$s dispersive readout of the
readout resonator, so that the probed transition stays saturated during the measurement.
Repetitions with the probe on and off are interleaved to cancel slow drifts. On resonance the probe rotates the readout signal without changing
its magnitude, so we record the phase of the probe-on signal relative to the probe-off one.

The spectrum (Fig.~\ref{sfig:thermometry}b) shows two peaks, which we assign to the transitions
$|\psi_0^\pm\rangle \leftrightarrow |\psi_1^\mp\rangle$ at $\omega_{01}$ and
$|\psi_1^\pm\rangle \leftrightarrow |\psi_2^\mp\rangle$ at $\omega_{12}$.
Applying a probe tone to the transition between manifolds $j$ and $k$ for longer than their relaxation and dephasing times equalises their populations, shifting the off-resonant readout phase by $D_{jk}$ (Ref.~\cite{adinolfi2025}, Eqs.~(S19) to (S23)). Assuming population conservation $p_0 + p_1 + p_2 = 1$, the ratio of the two fitted line heights determines the manifold populations $p_n$ of $|\psi_n^\pm\rangle$ via
\begin{equation}
\frac{D_{12}}{D_{01}} = \eta_{12} \frac{p_1 - p_2}{p_0 - p_1} \quad (\text{Ref. [9], Eq. S24}),
\end{equation}
where $\eta_{12}$ represents the step in readout response from $|\psi_1^\pm\rangle \to |\psi_2^\pm\rangle$ relative to that from $|\psi_0^\pm\rangle \to |\psi_1^\pm\rangle$. Here, we take $\eta_{12} = 1$.
Because the $|\psi_2^\pm\rangle \leftrightarrow |\psi_3^\mp\rangle$ transition line is not experimentally resolved, we bound the second-manifold population within $0 \le p_2 \le 2\%$, a range that encompasses the value of $(1.33 \pm 0.52)\%$ reported in Ref.~\cite{adinolfi2025}. Propagating the fit covariance matrix via Monte Carlo sampling yields $p_1 = 12.6^{+1.0}_{-1.1}\%$ for $p_2 = 0$ and $p_1 = 13.9 \pm 1.0\%$ for $p_2 = 2\%$. An oscillator at $\omega_a$ in thermal equilibrium with the same ratio $p_1/p_0$ would have an effective temperature of $93 \pm 5$~mK and $100 \pm 5$~mK for these two limits, respectively, corresponding to thermal occupations of $\bar{n}_{\mathrm{th}} = 0.17$ and $0.20$.

The populations rest on two assumptions. A smaller response step to the second manifold,
$\eta_{12} < 1$, would raise $p_1$. And the temperature attributes all of $p_1$ to thermal
excitation, whereas single-photon loss also populates $|\psi_1^\pm\rangle$ by quantum
heating, even at zero temperature~\cite{adinolfi2025}, so it bounds the temperature of the
oscillator's bath from above, and $\bar n_{\mathrm{th}}$ is an upper limit on the oscillator's thermal occupation $n_{\mathrm{th},a}$. Both mechanisms count in the heating rate $\Gamma_\uparrow$.
Since $T_1$ here exceeds $(\kappa_b/2)^{-1}$ more than sixfold, the cooling is inactive,
and $p_1/p_0 = 0.14$ to $0.16$ estimates the ratio $\Gamma_\uparrow/\Gamma_\downarrow$ of
Eq.~\eqref{eq:floor} at this drive, so the cooling competes with $\bar n_{\mathrm{th}}$ rather than $n_{\mathrm{th},a}$ (Eq.~\eqref{eq:floorcond}).

\section{The intra-well spectrum and its resonances}
\label{note:gap}

In this section, we derive the intra-well level spacing $\omega_{01}$ for arbitrary drive detuning and impose the resonance condition $\omega_{01}=\delta_b$ used in the main text. From this condition, we determine the lowest photon number at which the resonance can be reached at a fixed flux bias. We then map the resonance in the drive plane and compare it with the higher-order resonances allowed by the exchange interaction. Throughout the derivations below, $\Delta$ includes the Stark shift of Eq.~\eqref{eq:kcq} (Methods), $\Delta\to\Delta-\chi_{ab}|\beta|^2$, such that
$\delta_b+2\Delta=L-2\chi_{ab}|\beta|^2$. This correction is less than $1\%$ of $L$ at every flux bias in Fig.~\ref{sfig:minimum} and is included in the comparison of Sec.~\ref{note:gapmin}.

\subsection{Intra-well level spacing}

We start from the Kerr-cat Hamiltonian, Eq.~\eqref{eq:kcq} (Methods), and treat a single well.
With $K > 0$ the potential landscape bends downward as photon number increases. In the rotating frame, the stable classical Kerr-cat states at \(+\alpha\) and \(-\alpha\) therefore correspond locally to maxima of the quasienergy landscape, rather than minima. The intra-well transition appears in the laboratory frame at
$\omega_{\mathrm{sq}}/2 - \omega_{n-1,n}$, below the drive subharmonic. Displacing the
oscillator onto the classical fixed point of the well, $\hat a \to \alpha + \hat d$, removes
the term linear in $\hat d$ for the mean field
\begin{align}
  \alpha^2 &= \frac{\Delta + 2\varepsilon_2}{2K},
  \label{eq:nbar}
\end{align}
and leaves, up to a constant,
\begin{align}
  \hat H_{\mathrm{KCQ}}
  &= -\left(\Delta + 4\varepsilon_2\right)\hat d^\dagger\hat d
     - \frac{\Delta}{2}\left(\hat d^{2} + \hat d^{\dagger 2}\right) \nonumber\\
  &\quad - 2K\alpha\left(\hat d^\dagger\hat d^{2} + \hat d^{\dagger 2}\hat d\right)
     - K\hat d^{\dagger 2}\hat d^{2}.
  \label{eq:hdisplaced}
\end{align}
The quadratic part is diagonalised by the Bogoliubov rotation
$\hat d = \hat c\cosh r - \hat c^\dagger\sinh r$, with
\begin{align}
  \tanh 2r &= \frac{\Delta}{\Delta + 4\varepsilon_2},
  \label{eq:tanh2r}
\end{align}
which exists whenever the mean field does, $\Delta + 2\varepsilon_2 > 0$. The rotation
brings the quadratic part to $-\omega_{01}\hat c^\dagger\hat c$, with the intra-well level
spacing
\begin{align}
  \omega_{01} &= \sqrt{\left(\Delta + 4\varepsilon_2\right)^2 - \Delta^2}
  = 4K\alpha^2 e^{-2r}.
  \label{eq:omega01}
\end{align}
At $\Delta = 0$ the rotation is trivial and $\omega_{01} = 4\varepsilon_2 = 4K\alpha^2$. For
$\Delta > 0$ the well is squeezed along the axis joining the two wells, and the spacing
falls below $4K\alpha^2$ by the factor $e^{2r}$. The squeezing also adds photons: the ground
state of the well holds $\bar n = \langle\hat a^\dagger\hat a\rangle = \alpha^2 + \sinh^2 r$.

Equation~\eqref{eq:omega01} drops the cubic and quartic terms of
Eq.~\eqref{eq:hdisplaced}, and it runs high. At $\Delta = 0$ the coherent state
$|\alpha\rangle$ is an exact eigenstate, so the ground level does not shift, while the cubic
term couples the first and second excited levels with matrix element
$2\sqrt2K\alpha$ across a spacing $4K\alpha^2$; second-order perturbation theory therefore
lowers $\omega_{01}$ by $2K$. At finite detuning the ground state shifts as well and
partly cancels this, and to order $\alpha^{0}$
\begin{align}
  \omega_{01} &\approx 4K\alpha^2 e^{-2r} - K\left(1 - 2e^{-2r} + 3e^{-4r}\right).
  \label{eq:omega01corr}
\end{align}
For $\Delta \ge 0$ the correction lies between $2K/3$ and $2K$, so the relative error of
Eq.~\eqref{eq:omega01} falls as $\alpha^{-2}$. Against exact diagonalisation of
$\hat H_{\mathrm{KCQ}}$, Eq.~\eqref{eq:omega01corr} reduces that error by one to two orders
of magnitude. We use the harmonic form below because it gives closed expressions; the
two-mode fit diagonalises the full Hamiltonian.

In the rotated basis the one-excitation term of Eq.~\eqref{eq:vertices} (Methods) splits into
two processes,
\begin{align}
  2g_{21}\alpha\left(\hat d\,\hat b^\dagger + \hat d^\dagger\hat b\right)
  &= 2g_{21}\alpha\cosh r\left(\hat c\,\hat b^\dagger + \hat c^\dagger\hat b\right)
     \nonumber\\
  &\quad - 2g_{21}\alpha\sinh r\left(\hat c^\dagger\hat b^\dagger + \hat c\,\hat b\right).
  \label{eq:twoprocesses}
\end{align}
With the intra-well mode at $-\omega_{01}\hat c^\dagger\hat c$ and the readout resonator at
$-\delta_b\hat b^\dagger\hat b$, the first term, which converts an intra-well excitation
into a readout photon, is stationary at
\begin{align}
  \omega_{01} &= \delta_b,
  \label{eq:resonance}
\end{align}
and the second, which creates them in pairs, oscillates at $2\delta_b$ and averages out.
With the readout resonator above the squeezing drive, $\delta_b < 0$, the pair-creation
term would be the stationary one, and the same resonance would heat the well. The
conversion amplitude exceeds the coupling of Eq.~\eqref{eq:beamsplitter} of the main text
by $\cosh r$; Sec.~\ref{note:merging} gives the splitting it opens.

\subsection{Minimum photon number at a fixed flux bias}
\label{note:gapmin}

The squeezing frequency moves $\Delta$ and $\delta_b$ together, and their combination
$\delta_b + 2\Delta = L$ is set by the flux bias alone. At a fixed flux bias the resonance
condition, Eq.~\eqref{eq:resonance} with the spacing of Eq.~\eqref{eq:omega01}, therefore
fixes the photon number at the autoparametric resonance, $\bar n_{\mathrm{apr}}$, as a
function of $\Delta$ only. Equation~\eqref{eq:omega01} gives the mean field $\alpha^2$, which
near the minimum lies below $\bar n$ by less than half a photon (see below), so
\begin{align}
  \bar n_{\mathrm{apr}} &\approx \frac{\Delta + \sqrt{\Delta^2 + \delta_b^2}}{4K},
  \qquad \delta_b = L - 2\Delta .
  \label{eq:ncoll}
\end{align}
It is smallest at $\Delta = 3L/10$, $\delta_b = 2L/5$, and near that point
\begin{align}
  \bar n_{\mathrm{apr}} &\approx \frac{L}{5K} + \frac{\left(\Delta - 3L/10\right)^2}{KL}.
  \label{eq:nmin}
\end{align}
The minimum is shallow: missing the optimum by $L/10$ in $\Delta$ raises
$\bar n_{\mathrm{apr}}$ by about $5\%$. It lies on the ray $\Delta = 6\varepsilon_2$ of the
drive plane at every flux bias, so $\tanh 2r = 3/5$ there and the well is squeezed by
$r = \tfrac12\ln 2$, or $3$~dB, independently of $L$ and $K$. Two consequences follow:
$\omega_{01} = 2K\alpha^2$ at the minimum, and the ground state of the well holds
$\bar n = \alpha^2 + 1/8$ photons. The anharmonic correction of Eq.~\eqref{eq:omega01corr}
equals $3K/4$ on this ray and raises the minimum by $3/20$ of a photon at any $L$. The same
argument with $\omega_p$ in place of $\omega_b$ gives the minimum for the Purcell-filter
resonance.

\begin{figure}[tbp]
  \centering
  \includegraphics{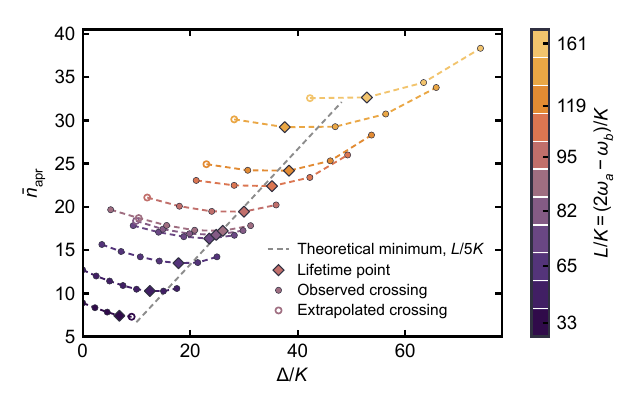}
  \caption{\textbf{Photon number at the readout resonance across flux biases.} Photon
  number $\bar n_{\mathrm{apr}}$ at the resonance $\omega_{01} = \delta_b$ against $\Delta/K$,
  for eleven flux biases coloured by $L/K$. Each point is one squeezing frequency, with
  $\bar n_{\mathrm{apr}}$ from the two-mode fit of its spectroscopy map
  (Sec.~\ref{note:twomodefit}) and $\Delta$ and $K$ the inputs of that fit. Dashed lines
  join the points of one flux bias. Filled circles are crossings observed in the map; open
  circles are crossings outside it, extrapolated from the fitted first excited level.
  Diamonds mark the points at which the lifetimes of Fig.~\ref{fig:flux} of the main text
  were measured. The grey dashed line is the theoretical minimum of Eq.~\eqref{eq:nmin},
  $L/(5K)$ at $\Delta = 3L/10$, traced as $L$ varies without the Stark shift, which would
  lower it by at most $0.22$ photons; the measured photon number lies above it
  (Sec.~\ref{note:gapmin}).}
  \label{sfig:minimum}
\end{figure}

Figure~\ref{sfig:minimum} compares Eq.~\eqref{eq:nmin} with the readout resonance measured
at eleven flux biases, $L/K = 33$ to $161$. At each bias the resonance was located at four
to eight squeezing frequencies, and its photon number $\bar n_{\mathrm{apr}}$ taken from the
two-mode fit. Where a spectroscopy map does not reach the avoided crossing, the fit of its
first excited level still predicts the drive at which $\omega_{01} = \delta_b$, and the
crossing is extrapolated; seven biases have one such point. The squeezing frequencies
sample the curve of Eq.~\eqref{eq:ncoll} coarsely, so the lowest point of each bias bounds
its minimum from above. That point exceeds $(L - 2\chi_{ab}|\beta|^2)/(5K)$ by $0.48$ to
$0.69$ photons, so the theoretical minimum describes the highest photon numbers to $2.1\%$ and
the lowest to $8.8\%$.

The offset has three parts. The fitted $\bar n$ (Sec.~\ref{note:twomodefit}) exceeds the
mean field $\alpha^2$ of Eq.~\eqref{eq:nbar} by $0.25$ to $0.42$ photons. The readout
resonator's admixture contributes $0.05$ to $0.12$ of this, and $0.21$ at the one bias where
$g_{21}$ sits at the edge of its search; the well alone contributes $0.20$ to $0.28$, of
which squeezing gives $1/8$ on the ray $\Delta = 6\varepsilon_2$ and the anharmonic terms of
Eq.~\eqref{eq:hdisplaced} about as much again. The distance of each lowest point from
$\Delta = 3L/10$ adds, by Eq.~\eqref{eq:nmin}, $0.23$ photons at the two highest biases,
where the point sits at $\Delta = 0.26L$, and at most $0.07$ elsewhere. The remaining $0.14$
to $0.19$ is close to the $3/20$ of the anharmonic correction. The readout-resonance
lifetimes of Fig.~\ref{fig:flux} of the main text were measured at or next to the lowest
point of each bias.

\subsection{Resonances in the drive plane}

The two-excitation term of Eq.~\eqref{eq:vertices} (Methods) rotates in the same way as the
one-excitation term. With
$\hat d^{2} = \hat c^{2}\cosh^2 r - \left(2\hat c^\dagger\hat c + 1\right)\sinh r\cosh r
+ \hat c^{\dagger 2}\sinh^2 r$,
\begin{align}
  g_{21}\left(\hat d^{2}\hat b^\dagger + \hat d^{\dagger 2}\hat b\right)
  &= g_{21}\cosh^2 r\left(\hat c^{2}\hat b^\dagger + \hat c^{\dagger 2}\hat b\right)
     - g_{21}\sinh 2r\left(\hat c^\dagger\hat c + \tfrac12\right)
       \left(\hat b + \hat b^\dagger\right) \nonumber\\
  &\quad + g_{21}\sinh^2 r\left(\hat c^{\dagger 2}\hat b^\dagger + \hat c^{2}\hat b\right).
  \label{eq:twoexcitation}
\end{align}
For $\delta_b > 0$ only the first part can be stationary. It converts two intra-well quanta
into one readout photon at $\omega_{02} = \delta_b$, and splits $|\psi_2, 0\rangle$ from
$|\psi_0, 1\rangle$ by $2\sqrt2\,g_{21}\cosh^2 r$, a factor $\cosh r/(\sqrt2\,\alpha)$ below
the splitting of the one-excitation term. The second part oscillates at $\delta_b$, and the
third, which creates two quanta together with a readout photon, is stationary only for
$\delta_b < 0$. Since $\omega_{02}$ is close to $2\omega_{01}$, the resonance needs about
half the spacing of Eq.~\eqref{eq:resonance}, and so a weaker drive. At a fixed flux bias,
the argument of Sec.~\ref{note:gapmin} with $\delta_b/2$ in place of $\delta_b$ puts its
minimum at $\Delta = 0$, at about $L/(8K)$ photons, against $L/(5K)$ for
$\omega_{01} = \delta_b$. The anharmonic terms of Eq.~\eqref{eq:hdisplaced}, which pull
$\omega_{02}$ below $2\omega_{01}$, raise it by about one photon.

The same link can heat the well. It joins the second rung to the ground level, so the
balance of Eq.~\eqref{eq:floor} (Sec.~\ref{note:floor}), written for it, compares
$n_{\mathrm{th},b}$ with the occupation that the oscillator's bath gives the second rung,
which is lower than that of the first. Its condition is therefore stricter than
Eq.~\eqref{eq:floorcond}, and a readout resonator that cools the first rung through the
one-excitation term can heat the second through the two-excitation term.

\begin{figure}[tbp]
  \centering
  \includegraphics[width=\textwidth]{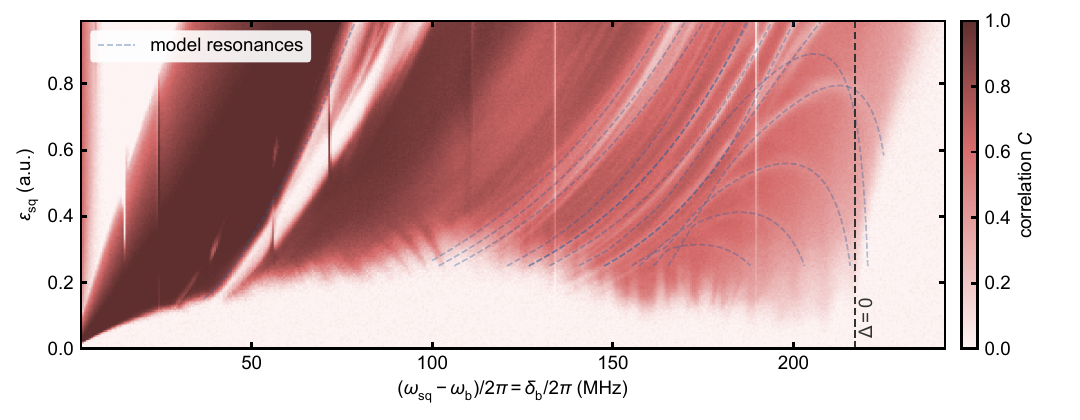}
  \caption{\textbf{Readout correlation across the drive plane.} Readout correlation $C$
  against the squeezing-readout detuning $\delta_b$ and the squeezing amplitude
  $\varepsilon_{\mathrm{sq}}$, at the flux bias of Fig.~\ref{fig:lifetime}a of the main
  text. Each pixel holds the cat at the operating point for $100~\mu$s and then correlates
  two back-to-back $4~\mu$s cat-quadrature readouts taken there,
  $C = \langle I_0 I_1 + Q_0 Q_1\rangle / \langle I_0^2 + Q_0^2\rangle$. Blue dashed lines show the computed resonance conditions,
  $\omega_{nm} = \delta_b$ of Eq.~\eqref{eq:higherorder} for $\delta_b/2\pi > 100$~MHz. They follow from the Kerr-cat Hamiltonian,
  Eq.~\eqref{eq:kcq} (Methods), with $K/2\pi = 1.97$~MHz. Lines for
  $\delta_b/2\pi > 100$~MHz come from a joint fit to the traced arcs, in which the delivered
  drive per unit $\varepsilon_{\mathrm{sq}}$ varies smoothly along $\delta_b$, compresses weakly at large
  drive, and shifts the oscillator frequency quadratically. The grey dashed
  line marks $\Delta = 0$, at $\delta_b/2\pi = 216.95$~MHz. The readout resonator is at
  $\omega_b/2\pi = 7.297$~GHz.}
  \label{sfig:corrmap}
\end{figure}

Figure~\ref{sfig:corrmap} maps the drive plane without the dynamical measurement of the
main text. The squeezing drive is held at the operating point for $100~\mu$s, and two
back-to-back $4~\mu$s cat-quadrature readouts, $16$~ns apart, are then taken at that same
point. Their correlation $C$ is close to one wherever the readout resolves the two wells and
the cat stays in its well through both readouts. Lines of low $C$ cross the region of high
correlation: along each, the correlation is lost within the $8~\mu$s of the two readouts.

The one-excitation resonances $\omega_{n-1,n} = \delta_b$ are marked in
Fig.~\ref{fig:lifetime}a of the main text. We do not assign the lines of
Fig.~\ref{sfig:corrmap} to particular levels. The exchange does allow further resonances:
the two-excitation term of Eq.~\eqref{eq:twoexcitation} is stationary where the spacing
between levels two apart equals $\delta_b$, and the anharmonic terms of
Eq.~\eqref{eq:hdisplaced}, which mix the levels, let the exchange connect levels further
apart. For comparison, Fig.~\ref{sfig:corrmap} overlays the conditions
\begin{align}
  \omega_{nm} &= \delta_b, \qquad n = 0, 1, 2, \quad m - n = 2 \text{ to } 6,
  \label{eq:higherorder}
\end{align}
with $\omega_{nm}$ the spacing between doublet $n$ and level $m$ of the ladder, and, below
$\delta_b/2\pi = 100$~MHz, $\omega_{01} = 2\delta_b$, a second-order process of the
exchange. A spacing inside the well grows with the drive and shrinks as $\Delta$ falls, so
over most of the map these conditions climb in $\varepsilon_{\mathrm{sq}}$ as $\delta_b$
grows, as most of the sloped lines do. The overlay follows several of the lines, but only
with the drive calibration and a frequency shift left free (Fig.~\ref{sfig:corrmap}), so it
does not identify the levels.

\section{Multilevel cooling at the autoparametric resonance}
\label{note:components}

In this section we show how the exchange cools the rungs of a well at the autoparametric resonance and how fast it drains them. In the shifted Fock basis, the exchange splits into three terms, which exchange zero, one, and
two intra-well excitations for one readout photon (Eq.~\eqref{eq:vertices}, Methods). At the
autoparametric resonance, $\omega_{01} = \delta_b$, the one-excitation term,
$2g_{21}\sqrt{\bar n}\,(\hat d\,\hat b^\dagger + \hat d^\dagger\hat b)$, is resonant.

\subsection{Multilevel anticrossings and the autoparametric band}
\label{note:chains}

\begin{figure}[tbp]
  \centering
  \includegraphics{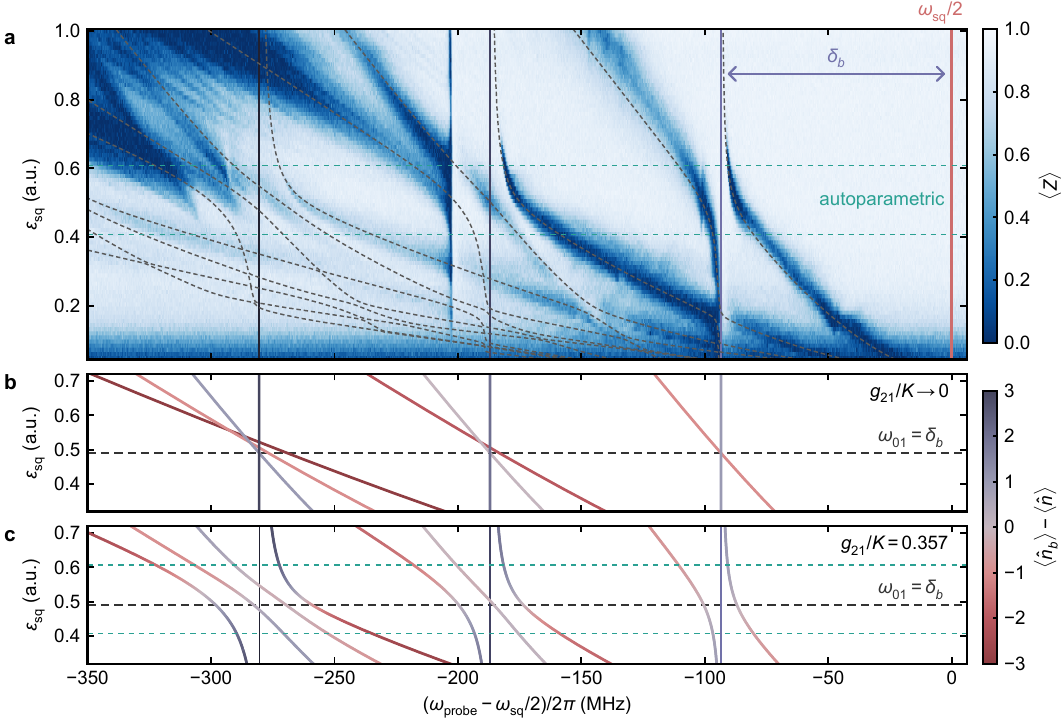}
  \caption{\textbf{Full spectrum of the multilevel anticrossings.}
    \textbf{a,} Excited-level spectroscopy of Fig.~\ref{fig:spectroscopy}a of the main text over its
  full range: expectation value $\langle Z\rangle$, against the squeezing
  amplitude $\varepsilon_{\mathrm{sq}}$ (a.u.) and the probe detuning from
  $\omega_{\mathrm{sq}}/2$. Grey dashed lines are the 13 lowest levels of the fitted two-mode model.
  Solid vertical lines mark $-N\delta_b$ for $N = 1$ to $3$, where the chain of index $N$ meets the
  readout resonator, and the red line marks $\omega_{\mathrm{sq}}/2$. The green dashed lines bound
  the autoparametric band of Fig.~\ref{fig:spectroscopy}a. The model follows the data at
  $-\delta_b$ and $-2\delta_b$; at $-3\delta_b$ it follows only part of them, and two branches lie
  $8$ to $20$~MHz from the nearest level. The vertical feature near $-203$~MHz is not a level of
  the model. Below $\varepsilon_{\mathrm{sq}} \approx 0.15$, $\langle Z\rangle$ is low even without the probe.
  \textbf{b,c,} Two-mode levels on the same axes, with the exchange off (\textbf{b},
  $g_{21}/K \to 0$, computed at $10^{-3}$) and at the $g_{21}/K = 0.357$ of this map's fit
  (\textbf{c}). Each line is coloured by $\langle \hat n_b\rangle - \langle \hat n\rangle$: red
  where the excitations are in the oscillator, purple where they are in the readout resonator, and
  pale where the two modes share them. Only the chains up to $N = 3$ are drawn. The vertical lines
  at $-N\delta_b$ are the states $|\psi_0, N\rangle$. The black dashed line marks
  $\omega_{01} = \delta_b$, at $\varepsilon_{\mathrm{sq}} = 0.49$ and $\bar n = 22.9$. Parameters: $\delta_b/2\pi = 93.5$~MHz,
  $\Delta/2\pi = 64.7$~MHz, $K/2\pi = 1.97$~MHz, $g_{21}/2\pi = 0.703$~MHz, and
  $\chi_{ab}/2\pi = 1.56$~kHz.}
  \label{sfig:resonances}
\end{figure}

\begin{figure}[tbp]
  \centering
  \includegraphics{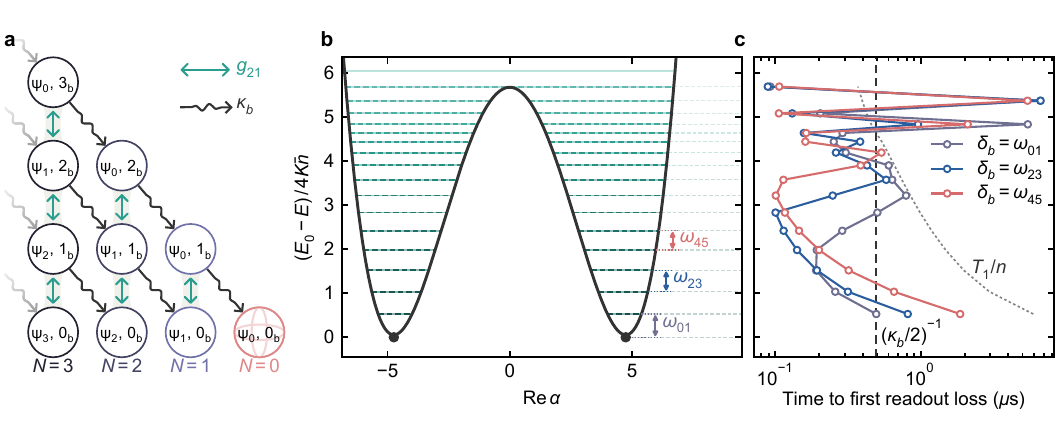}
  \caption{\textbf{Relaxation through the cascade.} \textbf{a,} Sites are the joint states
  $|\psi_n, n_b\rangle$. Green double arrows are the exchange $g_{21}$, which moves one rung down
  and one readout photon up and so stays in its column, a chain of constant $N = n + n_b$. Each
  chain meets the readout resonator at $-N\delta_b$ in Fig.~\ref{sfig:resonances}. Wavy arrows are
  readout loss at $\kappa_b$, the only process shown that changes $N$. Chains to $N = 3$ are drawn;
  the lattice continues.
  \textbf{b,} The metapotential and its levels at the simulated drive, $\varepsilon_2/K = 6.48$ and $\Delta/K = 32$, where the two parities are degenerate, with $\bar n = 22.5$. Arrows mark the three links of \textbf{c}.
  \textbf{c,} Simulated mean time from $|\psi_n, 0_b\rangle$ to the first readout photon loss, drawn at the energy of rung $n$ in \textbf{b}, with $\delta_b$ set to each link. The simulation follows the odd parity block and leaves out the oscillator's own decay. The dashed line is $(\kappa_b/2)^{-1}$; the dotted line, $T_1/n$ with $T_1 = 5.93~\mu$s, is a guide to the eye. Parameters: $K/2\pi = 1.97$~MHz, $g_{21}/2\pi = 0.703$~MHz, $\kappa_b/2\pi = 0.645$~MHz.}
  \label{sfig:cascade}
\end{figure}

The one-excitation term links $|\psi_n, n_b\rangle$ to $|\psi_{n-1}, n_b+1\rangle$. Here $|\psi_n, n_b\rangle$ is rung $n$ of a well with $n_b$ readout photons, the
well and parity labels of the main-text figures dropped where a statement holds for both. The states of the
joint system therefore organise into chains of constant $N = n + n_b$, and the chain of index $N$ meets the readout resonator in an $(N+1)$-level
anticrossing (Fig.~\ref{sfig:cascade}a). On the probe axis these anticrossings sit at $-N\delta_b$ from $\omega_{\mathrm{sq}}/2$,
near the drive where $\omega_{01} = \delta_b$; the anharmonicity of the ladder moves each one by its own amount. The $N = 1$
crossing at $-\delta_b$ is the two-level anticrossing of Fig.~\ref{fig:spectroscopy}a of the main text; the
$N = 2$ crossing at $-2\delta_b$, a three-level one, is visible in the same map one rung
higher in drive. The $N = 3$ crossing at $-3\delta_b$, a four-level one,
appears in the full range of that map (Fig.~\ref{sfig:resonances}a).

The coupling between $|\psi_n, n_b\rangle$ and $|\psi_{n-1}, n_b+1\rangle$ is
\begin{equation}
  V_{n,n_b} = \big(g_{21}\langle\psi_{n-1}|\hat a^{2}|\psi_n\rangle
  - \chi_{ab}\beta\,\langle\psi_{n-1}|\hat a^\dagger\hat a|\psi_n\rangle\big)\sqrt{n_b+1},
  \qquad \beta = \varepsilon_2/g_{21},
  \label{eq:link}
\end{equation}
where the second term is the residual cross-Kerr term of the readout-displaced frame, which lowers
the first link by $1.4\%$. The matrix elements grow along the ladder, and the readout resonator adds
the bosonic factor $\sqrt{n_b+1}$, so the later links of a chain are stronger than the first.

\label{note:merging}
The two-photon drive sets the intra-well ladder, so each link reaches its resonance
condition $\omega_{n-1,n} = \delta_b$ at its own squeezing amplitude. The ladder is
anharmonic, so the spacings differ from rung to rung by about $2K$. When the width of an
anticrossing is smaller than this spread, the rung resonances are
resolved: a single drive setting addresses one link, and the higher rungs are drained far
more slowly than the first. When it is larger, the anticrossings overlap: the
rungs merge into one band that shares the readout resonator's character, and one drive setting
drains the whole low ladder (Fig.~\ref{sfig:resonances}c). Our device is in this merged regime:
the width of the anticrossings is about $5$ times the spread for $n = 2$ to $5$, and the first
is about $21$ times $\kappa_b$. The drives at which the individual links $n \to n-1$ come through resonance
span the autoparametric band shaded in Fig.~\ref{fig:spectroscopy}a of the main text.

\subsection{Relaxation through the cascade}
\label{note:cascade}

When the links are resolved, an excitation in rung $n$ relaxes one link at a time, and each step
waits for a readout photon to decay. When several links are within their width of resonance, the
exchange moves the excitation down the ladder coherently, and the readout photons it creates then
decay at a rate proportional to their number. The last step of every path is the link
$|\psi_1,0\rangle \leftrightarrow |\psi_0,1\rangle$.

This last link limits the relaxation rate to $\kappa_b/2$, the strong-coupling limit. On resonance, the eigenstates of the
pair are equal superpositions of $|\psi_1,0\rangle$ and $|\psi_0,1\rangle$. Each holds half a
readout photon, so each decays at $\kappa_b/2$, however strong the exchange.

Two parameters bring transitions that are not exactly resonant closer to this rate. Increasing
$\kappa_b$ broadens the readout resonance, so a detuned transition couples through its Lorentzian
tail, but while the readout resonator can be adiabatically eliminated, the rate of the resonant
transition falls inversely with $\kappa_b$. Increasing $g_{21}$ instead hybridises the rungs into
one band, which raises the rates of the detuned rungs towards that of the resonant one. The readout
linewidth therefore sets how fast the ladder can drain, and the exchange sets how many transitions
reach that limit. The doublets of the lower rungs are closer to degenerate, so they contribute less
to tunnelling between the wells, and moving an excitation down the ladder generally helps.

To see how fast the ladder drains inside the autoparametric band of Fig.~\ref{fig:spectroscopy}a of the main text, we simulate the rung ladder coupled to the lossy readout resonator, without the oscillator's own decay, and compute the mean time from $|\psi_n, 0_b\rangle$ to the first readout photon loss (Fig.~\ref{sfig:cascade}c). The curve with $\delta_b = \omega_{01}$ is the device at its resonance, $\varepsilon_2/K = 6.48$. The other two curves keep this drive and lower $\delta_b$ to $\omega_{23}$ and $\omega_{45}$, which puts $\omega_{01}/\delta_b$ at $1.09$ and $1.21$, inside the band. They stand in for other positions in the band: the experiment moves through it by changing the drive at fixed $\delta_b$. With $\delta_b = \omega_{01}$ this time is close to
$(\kappa_b/2)^{-1}$ for $n = 1$. For rungs up to $10$ it is shorter, down to about $40\%$ of that, because the exchange spreads the
excitation into states with more readout photons, which are lost sooner. Moving the resonance to a
higher link moves the fastest rungs up and slows the first. Above rung $10$, near the saddle of the metapotential, the two parities stop forming doublets: one parity becomes a small orbit at the centre of phase space, with few photons, and the other a large orbit around both wells. In the odd block shown, the centre states are rungs $13$ and $15$. The exchange barely couples these to the rungs below, so the time is much longer, and the oscillator's own decay out of them is also slower than $T_1/n$.

\subsection{Thermal floor of the cooling}
\label{note:floor}

On resonance, excitations are exchanged coherently between the excited manifold and the readout resonator. Photon loss from the resonator provides cooling, while thermal photons drive the reverse process, setting a floor on the excited-state population. The cycle drawn in Fig.~\ref{fig:concept}b of the main
text is the net of the two directions.
Writing $\Gamma_\uparrow$ for the heating rate out of the code space, $\Gamma_\downarrow$ for the oscillator's intrinsic relaxation of the first excited manifold, $\Gamma_1$ for
the cooling rate of that manifold at zero readout occupation, and $n_{\mathrm{th},b}$ for the readout resonator's thermal occupation, the rate
equation for its population $p_1$ relative to the code space reads
\begin{equation}
  \frac{p_1}{p_0} = \frac{\Gamma_\uparrow + n_{\mathrm{th},b}\,\Gamma_1}
                         {\Gamma_\downarrow + (1+n_{\mathrm{th},b})\,\Gamma_1}
  \;\simeq\; \frac{n_{\mathrm{th},b}}{1+n_{\mathrm{th},b}}
  \qquad (\Gamma_1 \gg \Gamma_\uparrow, \Gamma_\downarrow),
  \label{eq:floor}
\end{equation}
against $\Gamma_\uparrow/\Gamma_\downarrow$ without the cooling. A larger $\Gamma_1$ speeds up the approach to this floor but does not lower it,
and the cooling lowers the excited population at all only while
\begin{equation}
  n_{\mathrm{th},b} < \bar n_{\mathrm{th}} ,
  \label{eq:floorcond}
\end{equation}
where $\Gamma_\uparrow/\Gamma_\downarrow = \bar n_{\mathrm{th}}/(1+\bar n_{\mathrm{th}})$ (Sec.~\ref{note:thermometry}),
that is, while the readout resonator is colder, in occupation, than the effective bath that heats the
oscillator. Since the readout resonator sits near twice the oscillator frequency, the same physical
temperature gives it a far smaller occupation, which is what makes the scheme viable
against a warm oscillator.

\section{Autoparametric resonance with the Purcell filter}
\label{note:purcell}

In this section we describe how the Purcell filter cools the oscillator by the same mechanism as
the readout resonator. The filter is a second lossy mode near $2\omega_a$, and the cubic
nonlinearity of the SNAIL couples the oscillator to it through an exchange of the same form as
$g_{21}$, with strength $g_{21}^{(p)} = 3g_3\lambda_p$ (Sec.~\ref{note:hamiltonian}). In the shifted Fock
basis this exchange takes the form of Eq.~\eqref{eq:vertices} (Methods), and its
one-excitation term is resonant at $\omega_{01} = \delta_p$, where
$\delta_p = \omega_{\mathrm{sq}} - \omega_p$.

To leading order, the results of Secs.~\ref{note:gap} and~\ref{note:components} involve the
readout resonator only through its detuning $\delta_b$, exchange $g_{21}$, linewidth
$\kappa_b$, and thermal occupation $n_{\mathrm{th},b}$, and each holds for the filter with
$\delta_p$, $g_{21}^{(p)}$, $\kappa_p$, and $n_{\mathrm{th},p}$ in their place. The minimum
photon number at a fixed flux bias is Eq.~\eqref{eq:nmin} with $2\omega_a - \omega_p$ in place
of $L$. The chain of index $N$ meets the filter at $-N\delta_p$ on the probe axis
(Sec.~\ref{note:chains}). The last link of the cascade, from the first rung into one filter
photon, limits the relaxation rate to $\kappa_p/2$ (Sec.~\ref{note:cascade}). The ratio
$p_1/p_0$ approaches the floor $n_{\mathrm{th},p}/(1+n_{\mathrm{th},p})$ of Eq.~\eqref{eq:floor}, and the
cooling lowers it only while $n_{\mathrm{th},p} < \bar n_{\mathrm{th}}$
(Eq.~\eqref{eq:floorcond}). Every term of the exchange acts as the identity on the logical
qubit, so the emission into the filter carries no which-well information either
(Sec.~\ref{note:phaseflip}).

\begin{figure}[tbp]
  \centering
  \includegraphics{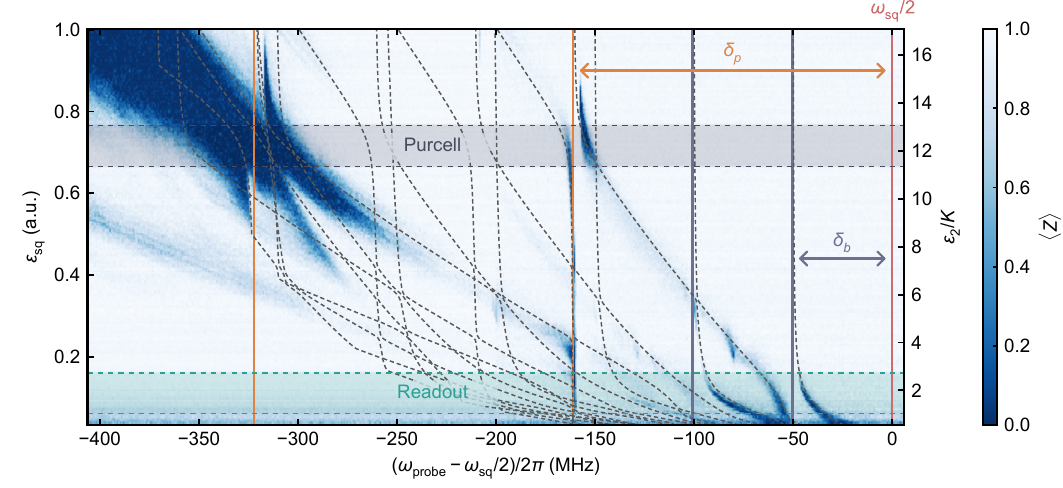}
  \caption{\textbf{Resonances with the readout resonator and the Purcell filter.}
  Excited-level spectroscopy, measured as in Fig.~\ref{sfig:resonances}a, at the flux bias and
  squeezing frequency of the blue diamond in Fig.~\ref{fig:lifetime}a of the main text ($K/2\pi = 1.97$~MHz). The colour is the expectation value $\langle Z\rangle$, which drops where the probe excites the oscillator. Grey dashed lines are the levels of the two-mode model of
  Sec.~\ref{note:twomodefit} with the Purcell filter added, fitted to the dips
  around the two resonances, which gives an exchange with the filter of
  $g_{21}^{(p)}/2\pi = 0.446$~MHz. Solid vertical lines mark where the transitions meet the
  readout resonator (purple) and
  the filter (orange), and the red
  line marks $\omega_{\mathrm{sq}}/2$. Arrows show $\delta_b/2\pi = 50.4$~MHz and $\delta_p/2\pi = 161.0$~MHz. Shaded bands mark the two autoparametric resonances. The map joins two
  runs at $-200$~MHz; the second used a stronger probe, which shifts the background.}
  \label{sfig:purcellmap}
\end{figure}

Figure~\ref{sfig:purcellmap} shows the two resonances in one map. It repeats the measurement
of Fig.~\ref{sfig:resonances}a at the flux bias of Fig.~\ref{fig:lifetime}a of the main text
and the squeezing frequency of its blue diamond, against the squeezing amplitude
$\varepsilon_{\mathrm{sq}}$ (a.u.) of the hold. The squeezing drive is ramped in $1.2~\mu$s
from the readout plateau to $\varepsilon_{\mathrm{sq}}$ and held there for $5~\mu$s with the
probe on, between two cat-quadrature readouts, and each row is repeated with the probe off. The
transitions meet the readout resonator at $-\delta_b$ and $-2\delta_b$, and the filter at
$-\delta_p$ and $-2\delta_p$. At $-2\delta_p$ the chain $N = 2$ forms the same three-level
anticrossing with the filter that it forms with the readout resonator at $-2\delta_b$ in
Fig.~\ref{sfig:resonances}a. The four-level anticrossings of the chain $N = 3$, which
Fig.~\ref{sfig:resonances}a shows for the readout resonator, do not appear in this map. The one
with the readout resonator, at $-3\delta_b$, needs a stronger probe than we used, and the one
with the filter, at $-3\delta_p$, lies outside the probe range. The filter lies below the
readout resonator (Table~\ref{stab:parameters}), so $\delta_p$ exceeds $\delta_b$ by
$\omega_b - \omega_p$ at every squeezing frequency, and as $\varepsilon_{\mathrm{sq}}$ rises,
$\omega_{01}$ reaches $\delta_b$ first and $\delta_p$ next, with the blue diamond between the
two. For the same reason, at any flux bias the minimum photon number of the filter resonance
exceeds that of the readout resonance by $(\omega_b - \omega_p)/5K$ (Eq.~\eqref{eq:nmin}).

The crucial difference between the two resonances is in how the qubit is read out. At the
readout resonance, the readout resonator hybridises with the intra-well excitations, which
degrades the cat-quadrature readout, so we read out at a higher drive with the dynamical
measurement (Methods). At the filter resonance, $\omega_{01}$ exceeds $\delta_b$ by the full
mode spacing $\omega_b - \omega_p$, so the exchange with the readout resonator is far from
resonance and the readout resonator stays unhybridised; in Fig.~\ref{sfig:purcellmap} its
anticrossings lie in the lower band, far below those of the filter. The filter then cools the
oscillator while the readout resonator reads it out, at the operating point itself.

\section{Phase-flip time under autoparametric cooling}
\label{note:phaseflip}

The autoparametric cooling leaves the phase-flip time at the single-photon-loss limit. In the shifted Fock basis $\hat a \approx \hat Z\otimes(\alpha + \hat d)$, so $\hat a^2 \approx \hat I\otimes(\alpha + \hat d)^2$, and the cooling term of the exchange has the form $2g_{21}\sqrt{\bar n}\;\hat I\otimes(\hat d\,\hat b^\dagger + \hat d^\dagger\hat b)$ of Eq.~\eqref{eq:beamsplitter}. It therefore leaves the qubit sector untouched, and the emitted photon carries no which-well information.

Only single-photon loss and gain change the oscillator parity, at the total rate $\kappa_a[(1+2n_{\mathrm{th},a})\bar n + n_{\mathrm{th},a}]$, which gives the loss limit of Eq.~\eqref{eq:tzlimit} (Methods). The exchange changes the oscillator photon number in pairs and the readout loss acts on the readout resonator alone, so both preserve parity. Single-photon jumps flip the parity in every manifold, so the cooling changes their rate only through $\bar n$, the photon number of the state.

Most phase flips occur in the ground manifold. In a round trip through an excited manifold, heating by $\hat a^\dagger$ flips the parity and native relaxation by $\hat a$ flips it back. The two-photon autoparametric cooling relaxes the oscillator with $\hat a^2$ instead, which preserves the parity, so the flip from the heating stays.

The opposite holds for heating through pure dephasing, with jump operator $\sqrt{\kappa_\phi}\,\hat a^\dagger\hat a$. Since $\hat a^\dagger\hat a \approx \hat I\otimes(\alpha + \hat d^\dagger)(\alpha + \hat d)$, this jump excites the intra-well mode and preserves parity. The native relaxation of that excitation is a single-photon jump, which flips the parity. The exchange removes the excitation with $\hat a^2$ and keeps the parity, so for this heating the cooling avoids a phase flip.

\end{document}